\documentclass[twocolumn]{aastex63}
\usepackage{amsmath}
\usepackage{bm}

\received{today}
\submitjournal{ApJ}

\shorttitle{The Effects of CR Diffusion and Radiative Cooling on the Galactic Wind from the MW}
\shortauthors{Shimoda \& Inutsuka}
\begin{document}

\title{The Effects of Cosmic-Ray Diffusion and Radiative Cooling on the Galactic Wind from the Milky Way}

\correspondingauthor{Jiro Shimoda}
\email{shimoda.jiro@k.mbox.nagoya-u.ac.jp}

\author[0000-0003-3383-2279]{Jiro Shimoda}
\affiliation{Department of Physics, Graduate School of Science, Nagoya University, \\
Furo-cho, Chikusa-ku, Nagoya 464-8602, Japan}

\author[0000-0003-4366-6518]{Shu-ichiro Inutsuka}
\affiliation{Department of Physics, Graduate School of Science, Nagoya University, \\
Furo-cho, Chikusa-ku, Nagoya 464-8602, Japan}

%\nocollaboration{1}



\begin{abstract}
The effects of cosmic-ray diffusion and radiative cooling
on the structure of the Galactic wind are studied using a steady state approximation.
It is known that realistic cooling processes suppress the wind from launching.
The effects of cosmic-ray diffusion are also supposed to be unfavorable for launching the wind.
Both of these effects have not been studied simultaneously in a steady-state approximation of the wind.
We find 327,254 solutions of the steady-state Galactic wind
and confirm that: the effect of 
cosmic-ray pressure depends on the Alfv{\'e}n Mach number, the mass flux
carried by the wind does not depend on the cosmic-ray pressure directly
(but depends on the thermal pressure), and
the typical conditions found in the Galaxy may correspond to the wind solution
that provides metal polluted matters at a height of $\sim300$~kpc from the disk.
\end{abstract}

%\keywords{editorials, notices --- miscellaneous --- catalogs --- surveys}
%\keywords{ISM: jets and outflows --- (ISM:) cosmic rays --- stars: formation --- Galaxy: evolution}


\section{Introduction}
\label{sec:intro}
Supernovae inject momenta, energies, and cosmic-rays (CRs)
into the interstellar medium (ISM). These drive the dynamics of
ISM~\citep[e.g.,][]{mckee77} and eventually result in the cloud formation~\citep{inutsuka15}.
The dynamics of ISM are
controlled by the pressures of thermal gas, turbulence, CRs, and
magnetic fields, which are comparable to each other
\citep{boulares90,ferriere01}.
The Galactic archaeological study shows that the overall star formation
rate in the Galaxy did not deviate much in the past $\sim8$~Gyr~\citep[e.g.,][]{haywood16}.
In this paper, we study this nonobvious star formation history in terms of
the mass budget of gaseous matter in the Galaxy.
\par
The star formation rate of the Milky Way (MW), $\sim1~{\rm M_\sun~yr^{-1}}$,
indicates a depletion of all gaseous matter in the Galactic disk with
a mass of $\sim10^9~{\rm M_\sun}$
within a time of $\sim1$~Gyr~\citep[e.g.,][]{kennicutt12}.
Therefore, to understand the star formation history
of the MW, which has been maintained during $\sim8$~Gyr with almost constant rate,
we must study the replenishment mechanisms of the gaseous matter.
Recent observations of metal absorption lines (e.g., \ion{Mg}{2}, \ion{O}{6}, etc.)
around external galaxies suggest that the circumgalactic medium (CGM) is huge mass
reservoir with a mass of $10^9\mathchar`-10^{12}~{\rm M_\sun}$~\citep[e.g.,][
and refernces therein]{tumlinson17}. Since these absorption lines are ubiquitously
observed around the host galaxy with a distance more than $\sim100$~kpc, we may naturally
consider the galactic wind as a metal transfer mechanism. Once the wind is really driven,
the metal-polluted CGM cools significantly by the radiative cooling and eventually falls
to the host galaxy. Thus, we have to study the possible conditions to launch such outflow
that is a part of the mass cycle, like a galactic fountain flow
\citep{shapiro76} but with a scale height of $100$~kpc.
\par
\citet{shapiro76} suggested the galactic fountain flow
based on observations of \ion{O}{6} absorption lines in our Galaxy.
Note that for the case of our galaxy, locations of gas responsible
for the absorption lines are still not constrained observationally~\citep[e.g.,][]{bregman07}.
Comparison the radiative cooling rate to the reheating rate of gas by supernovae,
the galactic fountain flow is considered to have a scale height of $\sim1$~kpc.
\citet{breitschwerdt91} considered that hot, tenuous gas coexisting with the cool,
condensed fountain flow is pushed by the CR pressure and eventually escapes from
the Galaxy as the Galactic wind. They showed steady-state solutions without
radiative cooling, diffusion of CRs, and other possible heating process.
\footnote{However, they studied the effects of Alfv{\'e}n wave damping using the model
of \citet{ipavich75}, which treats the Galactic system with a spherical geometry
acting as a point source of mass and energy at the center.
Their main analysis was done by using a model in cylindrical coordinates (the Galactic disk model)
without the wave damping.}
Radiative cooling was introduced by \citet{breitschwerdt99}, but the diffusion
of CRs was neglected. \citet{recchia16} introduced the CR diffusion and
heating due to the dissipation of Alfv{\'e}n waves, but they did not consider
the radiative cooling. Note that they solved the CR spectrum differing from
other studies and showed that the net CR pressure is almost the same as that calculated
by the fluid approximations.
Thus, we study steady-state outflow solutions including the radiative cooling,
CR diffusion, and heating due to the wave dissipation with fluid approximations.
\par
Recent numerical simulations study the dynamical role of CRs
in launching the Galactic wind~\citep[e.g.,][]{girichidis18,hopkins18}.
\citet{girichidis18} investigated outflows launched from the midplane
of the Galactic disk with solar neighborhood conditions
using a local box approximation with a size
of 0.5~kpc $\times$ 0.5~kpc $\times$ $\pm10$~kpc. They found that
the CR pressure can efficiently support the launching outflows
and strongly affects their phase structure. Their analysis was concentrated
at a height of $\la3$~kpc from the midplane.
\citet{hopkins18} performed a global simulation in the context of
galaxy formation and showed that the CR pressure can drive
the Galactic wind (outflow with a height of $\sim100$~kpc)
for the conditions of their simulated galaxy. In their simulation,
the inflow is also seen. These results may not be so surprising
in qualitatively because CRs do not lose their energy compared with
the thermal gas; the additional pressure can affect the thermal gas.
The aim of this paper is to show the effect of the CR pressure {\it explicitly}
by analyzing steady state-solutions of the outflow.
\par
This paper is organized as follows. In Sect.~\ref{sec:model},
we provide a physical model of the Galactic wind.
In Sect.~\ref{sec:wind equation}, the wind equation is analyzed,
and the role of CRs is discussed. The solutions of the Galactic
wind are shown in Sect.~\ref{sec:results}. Finally, we summarize
our results and discuss future prospects in Section~\ref{sec:discussion}.

\section{Physical Model}
\label{sec:model}
%
%%%%%%%%%%%%%%
\begin{figure}[htbp]
\centering
\includegraphics[scale=0.35]{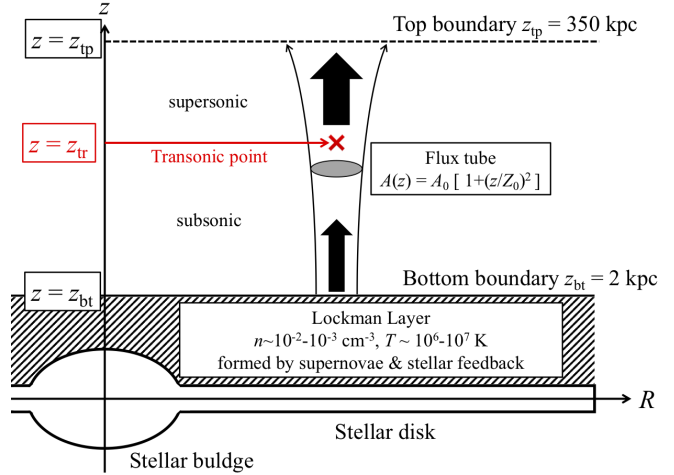}
\caption{
Schematic diagram of the outflow.
The Galactic disk, which consists of the stellar bulge and disk,
is approximated to be axially symmetric.
We solve the outflow from the bottom boundary $z_{\rm bt}=2$~kpc
to the top boundary $z_{\rm tp}=350$~kpc. The outflow is assumed
to travel along the flux tube $A(z)$, which is indicated by the
two thin black arrows. The red cross indicates the position of
the transonic point, $z=z_{\rm tr}$. We assume the existence
of hot and tenuous gas between the Galactic disk and bottom boundary.
This layer is called as the Lockman layer at which a number density
of $n\sim10^{-2}\mathchar`-10^{-3}~{\rm cm^{-3}}$ and temperature of
$10^6\mathchar`-10^7$~K are implied by the X-ray observations
~\citep[e.g.,][]{nakashima18} and numerical simulations
~\citep[e.g.,][]{girichidis18}.
}
\label{fig:schematic}
\end{figure}
%%%%%%%%%%%%%%
%
We study a steady outflow taking into account the effects of the CRs.
Figure~\ref{fig:schematic} shows the model geometry.
We approximate that the Galactic disk to be axially symmetric and work
in cylindrical coordinates. The radial distance of the disk is $R$, and the distance
perpendicular to the disk is $z$. 
There is  hot gaseous layer above the Galactic disk with a thickness of several kpc
that may be created due to supernova explosions
\citep[the so called Lockman layer;][]{lockman84,girichidis18}.
The outflow is solved from $z=2$~kpc
for the range of $1~{\rm kpc}\le R\le10~{\rm kpc}$ in this paper.  
\par
Plasma escaping from the galaxy may have a temperature comparable to
the virial temperature. Therefore,
we presume that the gas within the Lockman layer has a temperature comparable
to the virial temperature of the MW for a radius of $\sim100~{\rm kpc}$,
$T_{\rm vir}\sim3\times10^6~{\rm K}\left(M/10^{12}{\rm M_\sun}\right) \left(r/100~{\rm kpc}\right)^{-1}$,
and has a number density of $\sim10^{-3}~{\rm cm^{-3}}$.
The existence of gas with such temperature and density at $z=2~{\rm kpc}$ is implied by X-ray
observations~\citep[e.g.,][]{nakashima18} and numerical simulations~\citep[e.g.,][]{girichidis18}.
Supposing a magnetic field with a strength of $B\sim1~{\rm \mu G}$, we estimate
the plasma-$\beta$ as
$\beta\sim60
\left( \mu / 0.6 \right)
\left( n / 10^{-3}~{\rm cm^{-3}} \right)
\left( T / T_{\rm vir} \right)
\left( B / 1~{\rm \mu G} \right)^{-2}$,
where $\mu$, $n$, and $T$ are the mean molecular weight, number density,
and temperature, respectively. Thus, the dynamics of the outflow may not be
affected by the magnetic field that may be along the direction of travel of the outflow.
We assume that the magnetic field is always along with the flow in this paper.
\par
Since the geometry of the MW is that of a thin disk with assuming the axial symmetry,
the outflow `feels' the gravitational acceleration approximately along
the vertical direction $z$ at a relatively small height.
Thus, we treat a one-dimensional outflow
traveling along the $z$-direction
so that the required energy for launching the outflow becomes approximately minimum.
When the outflow reaches a height comparable
with the size of the Galactic disk, it feels a more multidimensional (i.e., spherical)
gravitational acceleration.
At such height, the outflow should cross a surface element like that of the spherical coordinate.
Hence, following~\citet{breitschwerdt91}, we assume that the outflow travels along
a cylindrical flux tube given by
%
%%%%%%%%%%%%%
\begin{eqnarray}
A(z) = A_0\left[ 1 + \left( \frac{ z }{ Z_0 } \right)^2 \right],
\end{eqnarray}
%%%%%%%%%%%%%
%
where $A_0$ is the area cross section of the flux tube at $z=0$, whose actual
value is not important in this paper.
The transition scale from vertical
to spherical is represented by $Z_0=15~{\rm kpc}$. Along this flux tube,
the divergence and gradient operations are represented by
%
%%%%%%%%%%%%%%
\begin{eqnarray}
{\rm div}  = \frac{1}{A(z)}
             \frac{ d }{ dz } A(z),~~~
{\rm grad} = \frac{ d }{ dz }.
\end{eqnarray}
%%%%%%%%%%%%%%
%
Thus, the mass and magnetic fluxes are given by
%
%%%%%%%%%%%%%%
\begin{eqnarray}
% mass flux
\label{eq:basic mass flux}
&& \frac{1}{A}\frac{ d }{ dz }
\left( A \rho v  \right) = 0,
\\
% div B
\label{eq:basic div B}
&& \frac{1}{A}\frac{ d }{ dz }
\left( A B \right) = 0,
\end{eqnarray}
%%%%%%%%%%%%%%
%
so that $\rho v A=$const. and $BA=$const. where $\rho$ and $v$ are
the mass density and velocity of the fluid, respectively. 
\par
The outflow mainly consists of nonrelativistic, thermal particles that are forced by
the pressures of the CRs and Alfv{\'e}n waves.
The basic equations (equation of motion and energy fluxes) of
this system can be given by~\citep[see][]{breitschwerdt91},
%
%%%%%%%%%%%%%%
\begin{eqnarray}
% equation of motion
\label{eq:basic equation of motion}
&& \rho v \frac{ dv    }{  dz }
= -       \frac{ d     }{  dz }\left( P_{\rm g}+P_{\rm cr}+P_{\rm w} \right)
  - \rho  \frac{ d\Phi }{  dz },
\\
\label{eq:basic energy flux of gas}
&& \frac{1}{A}\frac{ d }{ dz }
\left[ A
    \left\{
     \rho v
     \left( \frac{1}{2}v^2
          + \frac{ \gamma_{\rm g} }{ \gamma_{\rm g}-1 }\frac{ P_{\rm g} }{\rho}
    \right)
    \right\}
\right]
\nonumber \\
&& ~~~~~~~
=-       v \frac{ d     }{ dz }\left( P_{\rm cr}+P_{\rm w} \right)
 - \rho  v \frac{ d\Phi }{ dz }
 - n^2\Lambda + Q_{\rm w},
\\
% cosmic-ray transport
\label{eq:basic cosmic-ray transport}
&& \frac{1}{A}\frac{ d }{ dz }
   \left[A
        \left\{
         \frac{\gamma_c}{\gamma_c-1}\left(v+V_{\rm A}\right)P_{\rm cr}
        -\frac{\kappa}{\gamma_c-1}\frac{ d P_{\rm cr} }{ dz }
        \right\}
   \right]
\nonumber \\
&& ~~~~~~~
= \left( v+V_{\rm A} \right)\frac{ d P_{\rm cr} }{ dz },
\\
% wave energy
\label{eq:basic wave energy}
&& \frac{1}{A}\frac{ d }{ dz }
   \left[A\left( \frac{3}{2}v+V_{\rm A}\right)2P_{\rm w}\right]
%\nonumber \\
%&& ~~~~~~~
= v         \frac{ d P_{\rm  w} }{ dz }
- V_{\rm A} \frac{ d P_{\rm cr} }{ dz } - Q_{\rm w},
\nonumber \\
\end{eqnarray}
%%%%%%%%%%%%%%
%
where $P_{\rm g}$, $V_{\rm A}$,and $\gamma_{\rm g}=5/3$
are the pressure, Alfv{\'e}n speed,
and the adiabatic index of the thermal particles, respectively.
The gravitational acceleration due to the stars in the galaxy and
dark matter halo is $d\Phi/dz$, as given in the Sect.~\ref{sec:gravitational acceleration}.
The pressures of the CRs and Alfv{\'e}n waves, $P_{\rm cr}$ and $P_{\rm w}$,
are taken into account for the gas dynamics so that they appear in
the equation of motion~\eqref{eq:basic equation of motion}.
Equations~\eqref{eq:basic cosmic-ray transport} and \eqref{eq:basic wave energy}
describe the transport of the CRs and energy density of Alfv{\'e}n waves, respectively,
where $\kappa$ is the spatial diffusion coefficient of the CRs, and
$\gamma_c=4/3$ is the adiabatic index for relativistic particles.
These sets of equations were derived
previously, e.g., by~\citet{achterberg81a}.
The dissipation of Alfv{\'e}n waves is represented
by $Q_{\rm w}>0$, which results in the heating of the thermal particles (discussed in
the Sect.~\ref{sec:interplay}). $\Lambda$ is the radiative cooling function for the
thermal particles and is given in the Sect.~\ref{sec:radiative cooling}.
Here, we omit the hadronic losses of the CRs  that are inefficient in the tenuous
medium.\footnote{The collision time of the CRs via hadronic interactions can be estimated
as $\sim 1/n\sigma c
\sim100~{\rm Gyr}
(n/10^{-3}~{\rm cm^{-3}})^{-1}$
where $c$ and $\sigma\sim10^{-26}~{\rm cm^2}$
are the  speed of light and
the cross section of the hadronic interaction, respectively.}
\subsection{Gravitational Acceleration}
\label{sec:gravitational acceleration}
For the gravitational potential formed by the stars in the Galactic bulge and disk,
\citet{miyamoto75} gave a convenient pair of functions in the cylindrical coordinates as
%
%%%%%%%%%%%%%%
\begin{eqnarray}
\Phi_{\rm BD}(R,z)
= - \sum_{i=1}^2
    \frac{ G M_i }{ R^2+\left( a_i+\sqrt{z^2+b_i{}^2} \right)^2 },
\end{eqnarray}
%%%%%%%%%%%%%%
%
where the gravitational constant is $G$,
and the fit parameters are $a_i=(0;7.258)$~kpc,
$b_i=(0.495;0.520)$~kpc, and $M_i=(2.05\times10^{10};2.547\times10^{11})$~${\rm M_\sun}$ for
the bulge and disk, respectively.
\par
For the dark matter halo, we assume a Navarro-Frenk-White (NFW)-like density profile
\citep{nfw96} as
%
%%%%%%%%%%%%%%
\begin{eqnarray}
\varrho(x)=\frac{ \varrho_0 }{ x(1+x)^2 } - \frac{ \varrho_0 }{ x_v(1+x_v)^2 },
\end{eqnarray}
%%%%%%%%%%%%%%
%
where the dark matter density is $\varrho$, the normalized galactocentric radius
is $x = r/r_c$, and the parameters $\varrho_0$ and $x_v=r_v/r_c$ characterize the total
mass and extent of the dark matter halo, respectively.
The gravitational acceleration of the dark matter halo is obtained from the Poisson equation,
%
%%%%%%%%%%%%%%
\begin{eqnarray}
\frac{1}{r^2}\frac{ d              }{ dr }
\left(   r^2 \frac{ d\Phi_{\rm hl} }{ dr } \right)
=4\pi G\varrho,
\end{eqnarray}
%%%%%%%%%%%%%%
%
as
%
%%%%%%%%%%%%%%
\begin{eqnarray}
&& \frac{ d\Phi_{\rm hl} }{ dr }
=
\begin{cases}
%
%~~~~~~~~~~~~~~~~ for two column ~~~~~~~~~~~~~~~~~~~~~~
4\pi G\varrho_0 r_c
\left[ - \frac{ 1        }{ x(1+x)        }
       + \frac{ \ln(1+x) }{ x^2           }
\right. \\ \left.
~~~~~~~~~~~~
       - \frac{ x        }{ 3x_v(1+x_v)^2 } 
\right],~~~(x\le x_v)
%~~~~~~~~~~~~~~~~ for two column ~~~~~~~~~~~~~~~~~~~~~~
%
%~~~~~~~~~~~~~~~~ for one column ~~~~~~~~~~~~~~~~~~~~~~
%4\pi G\varrho_0 r_c
%\left[ - \frac{ 1        }{ x(1+x)        }
%       + \frac{ \ln(1+x) }{ x^2           }
%       - \frac{ x        }{ 3x_v(1+x_v)^2 }
%\right],~~~(x\le x_v)
%~~~~~~~~~~~~~~~~ for one column ~~~~~~~~~~~~~~~~~~~~~~
%
\\
\frac{ GM_{\rm hl} }{ r_c^2 x^2 },~~~(x>x_v)
\end{cases}
\end{eqnarray}
%%%%%%%%%%%%%%
%
where the total mass of the dark matter halo is
%
%%%%%%%%%%%%%%
\begin{eqnarray}
M_{\rm hl}
= 4\pi r_c{}^3
\left[   \ln( 1+x_v )
       - \frac{ x_v }{ 1+x_v }
       - \frac{ x_v{}^2 }{ 3(1+x_v)^2 }
\right].
\end{eqnarray}
%%%%%%%%%%%%%%
%
Setting the parameters as $\varrho_0=1.06\times10^7~{\rm M_{\sun}}~{\rm kpc^{-3}}$,
$M_{\rm hl}=10^{12}~{\rm M_{\sun}}$, and $r_v=300~{\rm kpc}$~\citep[e.g.][]{sofue12},
we obtain the core radius as $r_c=15.408$~kpc.
\par
\begin{figure}[htbp]
\centering
\includegraphics[scale=0.7]{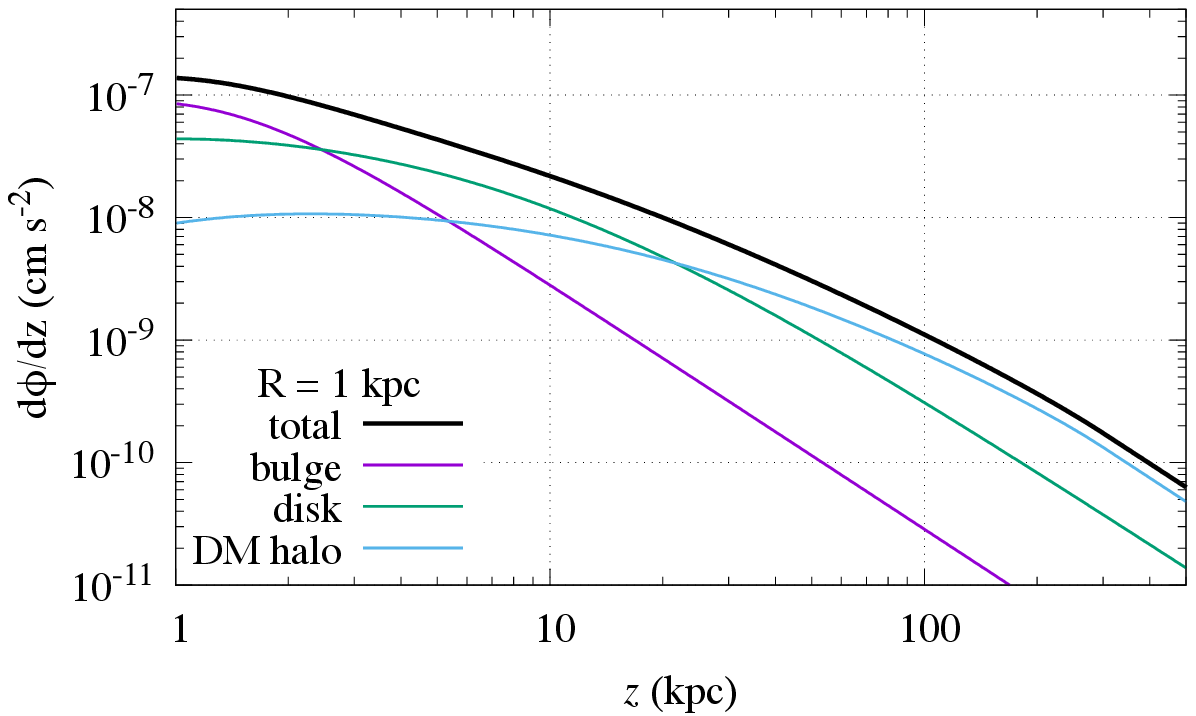}
\includegraphics[scale=0.7]{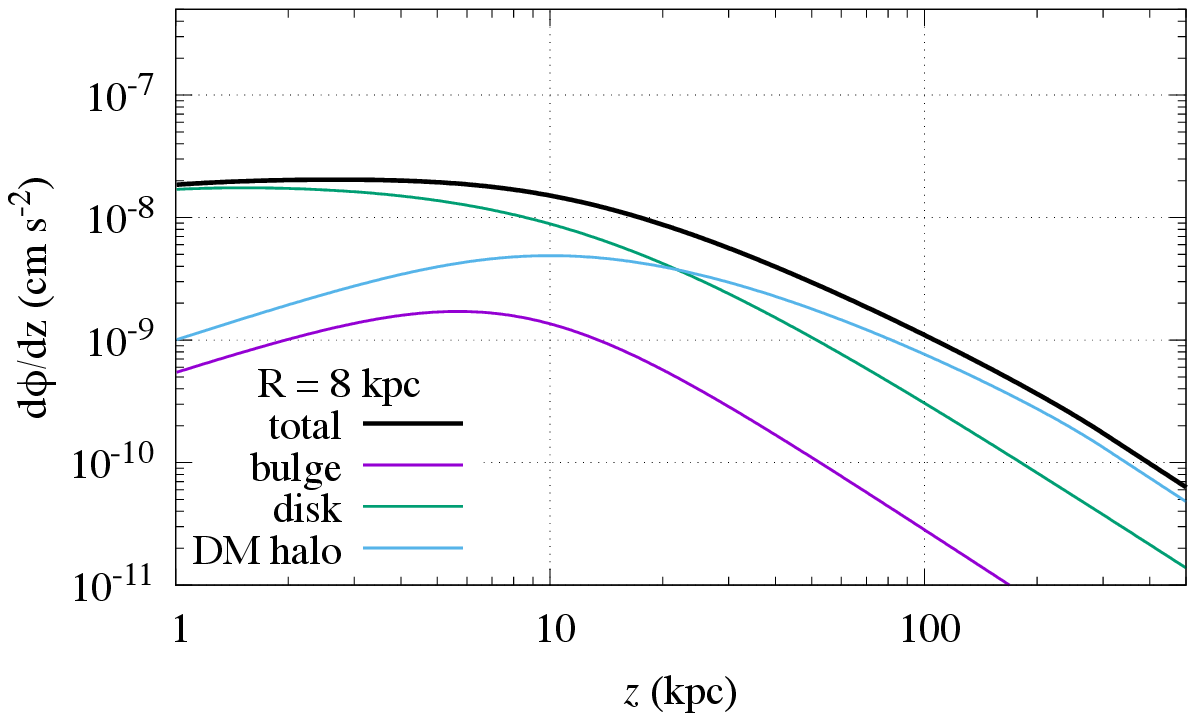}
\caption{The gravitational acceleration and its components as
a function of the vertical distance from the Galactic disk
at $R=1$~kpc (top panel) and $R=8$~kpc (bottom panel).
The black line shows the total acceleration. The purple and
green lines are the accelerations due to the bulge and disk stars,
respectively. The light blue line is the acceleration due to the
dark matter halo with the NFW profile.}
\label{fig:geff}
\end{figure}

The total gravitational acceleration is written as
%
%%%%%%%%%%%%%%
\begin{eqnarray}
  \frac{ d\Phi          }{ dz         }
= \frac{ d\Phi_{\rm BD} }{ dz         }
+ \frac{ z              }{ \sqrt{ R^2+z^2 } }
  \frac{ d\Phi_{\rm hl} }{ dr },
\end{eqnarray}
%%%%%%%%%%%%%%
%
and is represented by Figure~\ref{fig:geff} for $R=1~{\rm kpc}$ (top panel)
and $R=8~{\rm kpc}$ (bottom panel).
%
%%%%%%%%%%%%%%
%%%%%%%%%%%%%%
%
%
\subsection{Interplay between the Cosmic Rays and Alfv{\'e}n Waves}
\label{sec:interplay}
The CRs can excite the Alfv{\'e}n waves in the background plasma~\citep{lerche66,lerche67,
wetzel68,kulsrud69}. In the fluid approximation of CRs, the generation rate of the wave
can be estimated
as $-V_{\rm A}dP_{\rm cr}/dz$~\citep[e.g.,][]{kulsrud05} and appears on the right-hand
side of Eqs.~\eqref{eq:basic cosmic-ray transport} and \eqref{eq:basic wave energy} as the
energy sink and source term, respectively. Note that $dP_{\rm cr}/dz < 0$.
\par
On the other hand, the Alfv{\'e}n waves in a high $\beta$-plasma can be dissipated via (at least)
the nonlinear Landau damping~\citep{lee73,achterberg81b}. We assume a local equilibrium between the
wave generation and dissipation so that the energy sink term of the wave energy in Eqs.~\eqref{eq:basic
wave energy}, $-Q_{\rm w}$, is equal to $V_{\rm A}dP_{\rm cr}/dz$~\citep{volk81,zirakashvili96}.
The dissipated energy is converted to the thermal energy of gas; therefore $Q_{\rm w}$ appears on the
right-hand side of Eq.~\eqref{eq:basic energy flux of gas} as the heating term.
\par
CRs are scattered by the Alfv{\'e}n waves, which results in the diffusion of the CRs. The actual diffusion
coefficient of the CRs currently remains to be uncertain. We assume one of the most frequently invoked
coefficients that is estimated from the observations of CR compositions~\citep[see][for a recent
review]{hayakawa58,ginzburg64,gabici19},
%
%%%%%%%%%%%%%%
\begin{eqnarray}
\kappa = 3.3\times10^{28}~{\rm cm^2~s^{-1}}
\left( \frac{ P_{\rm M}/P_{\rm w} }{ 10^6          } \right)
\left( \frac{ B                   }{ {\rm 1~\mu G} } \right)^{-1},
\end{eqnarray}
%%%%%%%%%%%%%%
%
where $P_{\rm M}=B^2/8\pi$. For the dependence of $\kappa$, we adopt the case of pitch-angle
scattering due to small-amplitude Alfv{\'e}n waves~\citep{jokipii66}.
The energy density of the Galactic CRs may be mostly deposited by CRs with energies
of $1\mathchar`-10$~GeV~\citep[e.g.,][]{acero16}. Since an Alfv{\'e}n wave generated by a CR
has a wavelength comparable to the associated CR Larmor radius, the wave pressure $P_{\rm w}$ is
almost given by the field disturbance with a wavelength of $\sim10^{13}~{\rm cm}\left( E_{\rm cr}
/10~{\rm GeV} \right)\left( B/1~{\rm \mu G} \right)^{-1}$. Pitch-angle scattering becomes
strong when the CR interacts with the wave that has a comparable wavelength with the CR's Larmor radius.
Thus, we regard that the momentum averaged diffusion coefficient $\kappa$ is represented by
the coefficient of the $\sim$GeV CRs.

\subsection{Radiative Cooling}
\label{sec:radiative cooling}
Radiative cooling rates generally depend on the temperature via the ionization balance
of gas. In our case, the gas is tenuous, and its temperature
may be around the virial temperature of the MW as $T_{\rm vir}\sim
3\times10^6~{\rm K}\left(M/10^{12}{\rm M_\sun}\right) \left(r/100~{\rm kpc}\right)^{-1}$.
In such situation, atomic line emissions are the most important processes.
\par
For calculations of the radiative cooling function, we approximate that the
gas is in the optically thin limit for simplicity. Hence, the atomic
ionization state (level population) is given by the ratio of the
collisional ionization rate (excitation rate) to the recombination
rate (spontaneous transition rate). In this paper, we omit the 
photoionization and the charge-exchange reaction.\footnote{
\citet{gnat17} calculated the time-dependent cooling function
including the photoionization by the metagalactic radiation
field for the present time. The function is similar to
the pure collisional case for a range of number density,
$\ga10^{-5}~{\rm cm^{-3}}$.}
We calculate the bound-bound, free-bound, free-free, and the two-photon decays.
\par
The radiation power of line emission due to the transition from the upper level $u$
to the lower level $l$ (bound-bound) is estimated as~\citep[e.g.,][]{Osterbrock06}
%
%%%%%%%%%%%%%%
\begin{eqnarray}
P_{ul} = n_{\rm e}n_{\rm ion} E_{ul}q_{lu},
\end{eqnarray}
%%%%%%%%%%%%%%
%
where $n_{\rm e}$ and $n_{\rm ion}$ are the number densities of the electron and atom,
respectively. The emitted photon energy is the subtraction of the upper energy level $E_u$
and the lower energy level $E_l$, $E_{ul}=E_u-E_l$. The collisional excitation rate
[${\rm cm^3~s^{-1}}$] is given by
%
%%%%%%%%%%%%%%
\begin{eqnarray}
&& q_{lu} = 8.629\times10^{-6}
          \frac{ \Omega_{lu} }{ g_l }
          \frac{ {\rm e}^{-\frac{ E_{ul} }{ kT } } }{ \sqrt{T} },
\end{eqnarray}
%%%%%%%%%%%%%%
%
where $g_l$ is the statistical weight of the lower level,
$k$ is Boltzmann constant, and $T$ is the temperature of gas.
The collision strength is 
%
%%%%%%%%%%%%%%
\begin{eqnarray}
&& \Omega_{lu}
          = \frac{ 8\pi }{ \sqrt{3} }
            \frac{ g_l f_{lu} }{ E_{ul,{\rm Ryd}} } 
            \bar{g}(T),
\end{eqnarray}
%%%%%%%%%%%%%%
%
where $f_{lu}$ is the oscillator strength, and $E_{ul,{\rm Ryd}}$ is the photon
energy given in the Rydberg unit. The averaged Gaunt factor is $\bar{g}$. The value of the averaged
Gaunt factor is around unity, and determines the detailed temperature dependence
of the excitation rate. The precise data of the excitation rate (or $\bar{g}$)
are, however, still not available. In this paper, we use the following fitting function
\citep{mewe72}
%
%%%%%%%%%%%%%%
\begin{eqnarray}
\bar{g}(T) = 0.15
+ 0.28
\left[ \log \left( \frac{ \chi+1 }{\chi} \right)
     - \frac{ 0.4 }{ (1+\chi)^2 }
\right],
\end{eqnarray}
%%%%%%%%%%%%%%
%
where $\chi = E_{ul}/kT$, for the neutral atoms, while we assume $\bar{g}=1$ for the
ionized atoms. Note that the cooling function mainly depends on the ionization balance
of atoms rather than $\bar{g}$. For the oscillator strength and energy levels, we use
the data table given by National Institute of Standards and Technology.
For the calculation of the cooling function, it is sufficient to consider only
the allowed transitions from the ground state.
\par
For the calculations of continuum components, we follow \citet[][free-bound]{mewe86}
and \citet[][free-free and two-photon decays]{gronen78}. Then, integrating for the
photon frequency, we obtain the net radiation power and thus the cooling function.
\par
We solve for the 10 most abundant elements H, He, C, N, O, Ne, Mg, Si, S, and Fe
\citep{asplund09}. The ionization cross sections are given by \citet{janev93}
for H, and \citet{lennon88} for the others. The fitting functions for those data
are given by the International Atomic Energy Agency. We summarize the literatures for
the recombination rates in Table~\ref{tab:recombination}. We fit those data by the
Chebyshev polynomials with twenty terms. For the hydrogen-like atoms, we follow the fitting
function given by \citet{kotelnikov19}.
%
%%%%%%%%%%%%%%
%\begin{longrotatetable}
\begin{table*}
\tabletypesize{\scriptsize}
\centering
\caption{Literatures for the recombination rates. The superscript $^*$
denotes that we use the Mewe's formula for the radiative recombination
\citep{mewe80a,mewe80b}.}
\label{tab:recombination}
\begin{tabular}{cc cc cc}
\hline
Ion & Literature & Ion & Literature & Ion & Literature \\
\hline
 C$^{+1 }$ & \citet{nahar99}          &  Mg$^{+5 }$ & \citet{arnaud85}         &   S$^{+12}$ & \citet{mewe80a,mewe80b}  \\
 C$^{+2 }$ & \citet{nahar99}          &  Mg$^{+6 }$ & \citet{zatsarinny04}     &   S$^{+13}$ & \citet{mewe80a,mewe80b}  \\
 C$^{+3 }$ & \citet{nahar97}          &  Mg$^{+7 }$ & \citet{nahar95}          &   S$^{+14}$ & \citet{arnaud85}         \\
 C$^{+4 }$ & \citet{nahar97}          &  Mg$^{+8 }$ & \citet{arnaud85}         &   S$^{+15}$ & \citet{arnaud85}         \\   
 C$^{+5 }$ & \citet{nahar97}          &  Mg$^{+9 }$ & \citet{arnaud85}         &  Fe$^{+1 }$ & \citet{nahar97}          \\  
 N$^{+1 }$ & \citet{zatsarinny04}     &  Mg$^{+10}$ & \citet{arnaud85}         &  Fe$^{+2 }$ & \citet{nahar97}          \\  
 N$^{+2 }$ & \citet{nahar97}          &  Mg$^{+11}$ & \citet{arnaud85}         &  Fe$^{+3 }$ & \citet{nahar97}          \\
 N$^{+3 }$ & \citet{nahar97}          &  Si$^{+1 }$ & \citet{nahar00}          &  Fe$^{+4 }$ & \citet{nahar98}          \\
 N$^{+4 }$ & \citet{nahar97}          &  Si$^{+2 }$ & \citet{altun07}          &  Fe$^{+5 }$ & \citet{nahar99}          \\
 N$^{+5 }$ & \citet{nahar06}          &  Si$^{+3 }$ & \citet{mewe80a,mewe80b}  &  Fe$^{+6 }$ & \citet{arnaud85}         \\
 N$^{+6 }$ & \citet{nahar06}          &  Si$^{+4 }$ & \citet{zatsarinny03}     &  Fe$^{+7 }$ & \citet{nahar00}          \\
 O$^{+1 }$ & \citet{nahar98}          &  Si$^{+5 }$ & \citet{zatsarinny06}$^*$ &  Fe$^{+8 }$ & \citet{arnaud85}         \\
 O$^{+2 }$ & \citet{zatsarinny04}     &  Si$^{+6 }$ & \citet{zatsarinny03}     &  Fe$^{+9 }$ & \citet{arnaud85}         \\
 O$^{+3 }$ & \citet{nahar98}          &  Si$^{+7 }$ & \citet{mitnik04}$^*$     &  Fe$^{+10}$ & \citet{lestinsky09}$^*$  \\
 O$^{+4 }$ & \citet{nahar98}          &  Si$^{+8 }$ & \citet{zatsarinny04}     &  Fe$^{+11}$ & \citet{novotny12}$^*$    \\
 O$^{+5 }$ & \citet{nahar98}          &  Si$^{+9 }$ & \citet{nahar95}          &  Fe$^{+12}$ & \citet{hahn14}$^*$       \\
 O$^{+6 }$ & \citet{nahar98}          &  Si$^{+10}$ & \citet{arnaud85}         &  Fe$^{+13}$ & \citet{arnaud85}         \\
 O$^{+7 }$ & \citet{nahar98}          &  Si$^{+11}$ & \citet{arnaud85}         &  Fe$^{+14}$ & \citet{altun07}$^*$      \\
Ne$^{+1 }$ & \citet{arnaud85}         &  Si$^{+12}$ & \citet{arnaud85}         &  Fe$^{+15}$ & \citet{murakami06}$^*$   \\
Ne$^{+2 }$ & \citet{zatsarinny03}     &  Si$^{+13}$ & \citet{arnaud85}         &  Fe$^{+16}$ & \citet{zatsarinny04}     \\
Ne$^{+3 }$ & \citet{mitnik04}$^*$     &   S$^{+1 }$ & \citet{mewe80a,mewe80b}  &  Fe$^{+17}$ & \citet{arnaud85}         \\
Ne$^{+4 }$ & \citet{zatsarinny04}     &   S$^{+2 }$ & \citet{nahar95}          &  Fe$^{+18}$ & \citet{zatsarinny03}     \\
Ne$^{+5 }$ & \citet{nahar95}          &   S$^{+3 }$ & \citet{nahar00}          &  Fe$^{+19}$ & \citet{savin02}$^*$      \\
Ne$^{+6 }$ & \citet{arnaud85}         &   S$^{+4 }$ & \citet{altun07}          &  Fe$^{+20}$ & \citet{zatsarinny04}     \\
Ne$^{+7 }$ & \citet{arnaud85}         &   S$^{+5 }$ & \citet{arnaud85}         &  Fe$^{+21}$ & \citet{arnaud85}         \\
Ne$^{+8 }$ & \citet{nahar06}          &   S$^{+6 }$ & \citet{zatsarinny04}     &  Fe$^{+22}$ & \citet{arnaud85}         \\
Ne$^{+9 }$ & \citet{nahar06}          &   S$^{+7 }$ & \citet{zatsarinny06}$^*$ &  Fe$^{+23}$ & \citet{mewe80a,mewe80b}  \\
Mg$^{+1 }$ & \citet{mewe80a,mewe80b}  &   S$^{+8 }$ & \citet{zatsarinny03}     &  Fe$^{+24}$ & \citet{nahar01}          \\
Mg$^{+2 }$ & \citet{zatsarinny04}     &   S$^{+9 }$ & \citet{mitnik04}$^*$     &  Fe$^{+25}$ & \citet{nahar01}          \\
Mg$^{+3 }$ & \citet{arnaud85}         &   S$^{+10}$ & \citet{zatsarinny04}     &             &                          \\
Mg$^{+4 }$ & \citet{zatsarinny04}     &   S$^{+11}$ & \citet{nahar95}          &             &                          \\
\hline                                
\end{tabular}                         
\end{table*}
%\end{longrotatetable}
%%%%%%%%%%%%%%
%
\par
%
%%%%%%%%%%%%%%
\begin{figure}[htbp]
\centering
\includegraphics[scale=0.65]{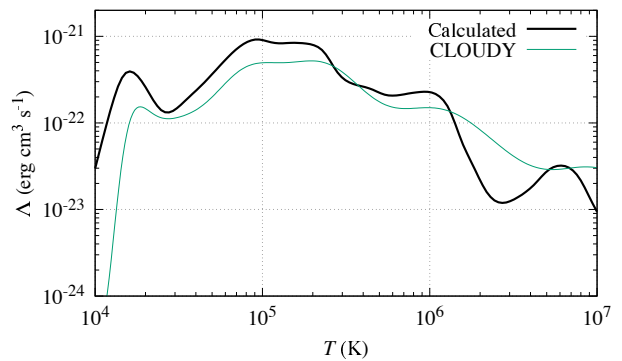}
\caption{Calculated cooling function $\Lambda$ (black).
We also show the cooling function calculated by the Cloudy
\citep[green,][]{ferland17} for comparison.}
\label{fig:cooling function}
\end{figure}
%%%%%%%%%%%%%%
%
Figure~\ref{fig:cooling function} shows the calculated cooling function $\Lambda$
in the collisional ionization equilibrium (black line) that is consistent with the cooling
function given by Cloudy~\citep[green,][]{ferland17}. Since we omit the charge-exchange reaction, which works
at $T\sim10^4$~K, the function is overestimated due to the survived H and lower ionized ions.
Interestingly, our $\Lambda$ shows a depression at $T\sim3\times10^6$~K compared with
the function given by Cloudy, around which the gas can be thermally stable ($d\ln\Lambda/d\ln T\ga2$).
This may be due to the updated recombination rates of Fe. In this paper, we concentrate to study the nature of
Galactic wind, deferring a further analysis on the depression for future work.
In this article, we assume the collisional ionization equilibrium and use this cooling function.
\par
Compton heating and photoionization heating are competitive processes against the radiative
cooling in general. We neglect them for simplicity. In the case of the MW (irradiated by the metagalactic
radiation field), they do not dominate over the cooling around the virial temperature unless $n\ll
10^{-5}~{\rm cm^{-3}}$~\citep[e.g.,][]{gnat17}.
The outflow with a mass transfer rate of $\sim\rho v R^2\sim0.2~M_\sun~{\rm yr^{-1}}
\left(n/10^{-3}~{\rm cm^{-3}}\right)
\left(v/100~{\rm km~s^{-1}}\right)
\left(R/10~{\rm kpc}\right)^2$ that is comparable to the star formation rate of the MW
has a number density of $n\sim10^{-3}~{\rm cm^{-3}}\gg10^{-5}~{\rm cm^{-3}}$.
When the number density reaches at $10^{-5}~{\rm cm^{-3}}$, the heating rate dominates
over the cooling rate at $T\la2\times10^{4}~{\rm K}$ with the solar metallicity and metagalactic
radiation field at the current time~\citep{gnat17}. The cooling rate becomes comparable to
that in the case of collisional ionization equilibrium (CIE) at $T\ga5\times10^4$~K. At a higher temperature, the cooling rate is smaller
than the CIE case by a factor of $2$ due to the photoionization yielding highly ionized ions.
Note that, at a given electron temperature, an increment of in highly ionized ions reduces the line intensity
because a potential energy of the bound electrons becomes large.
Thus, the radiative heating is not expected to be important, and we neglect it for simplicity.
\par
Heating due to the dissipation of the Alfv{\'e}n waves can be comparable to the
radiative cooling. Defining a scale height of the CR pressure as $H_{\rm cr}\equiv P_{\rm cr}|dP_{\rm cr}/dz|^{-1}$,
we estimate the ratio of the radiative cooling rate to the heating rate due to the wave dissipation as
%
%%%%%%%%%%%%%%
\begin{eqnarray}
&&
\frac{ n^2\Lambda }{ Q_{\rm w} }
\simeq0.91
\left( \frac{ n          }{ 10^{-3}~{\rm cm^{-3}}          } \right)^{5/2}
\left( \frac{ B          }{ 1~{\rm \mu G}                  } \right)^{-1}
\left( \frac{ P_{\rm cr} }{ 0.3~{\rm eV~cm^{-3}}            } \right)^{-1}
\nonumber \\
&& ~~~~~~~
\times
\left( \frac{ H_{\rm cr} }{ 10~{\rm kpc}                   } \right)
\left( \frac{ \Lambda    }{ 10^{-22}~{\rm erg~cm^3~s^{-1}} } \right).
\label{eq:cool to heat}
\end{eqnarray}
%%%%%%%%%%%%%%
%
Since the transport of the CRs is determined by a combination of the advection and diffusion
as described by Eq.~\eqref{eq:basic cosmic-ray transport}, the characteristic length
scale can be estimated as $H_{\rm cr}\sim\kappa/v\sim10~{\rm kpc}~(\kappa/3\times10^{28}~{\rm cm^2~s^{-1}})
(v/10~{\rm km~s^{-1}})^{-1}$.
Thus, the gas heating process may be mainly determined by the dissipation of Alfv{\'e}n waves
rather than the radiative process.
\section{wind equation and transonic point analysis}
\label{sec:wind equation}
From the basic equations~\eqref{eq:basic equation of motion}-\eqref{eq:basic wave energy},
we can derive the wind equation as
%
%%%%%%%%%%%%%%
\begin{eqnarray}
\frac{v'}{v}
=\left( \frac{A'}{A} \right)
 \frac{ \left( C_{\rm g}{}^2 + C_{\rm w}{}^2 \right) - V_{\rm g}{}^2 }
      { v^2 - \left( C_{\rm g}{}^2 + C_{\rm w}{}^2 \right) },
\label{eq:wind equation}
\end{eqnarray}
%%%%%%%%%%%%%%
%
where the prime indicates the derivative with respect to $z$ (e.g., $v'=dv/dz$),
and we define
%
%%%%%%%%%%%%%%
\begin{eqnarray}
\label{eq:Cg}
&& C_{\rm g}{}^2 \equiv \frac{ \gamma_{\rm g}P_{\rm g} }{\rho},
\\
\label{eq:Cw}
&& C_{\rm w}{}^2 \equiv \frac{ 3M_{\rm A}+1 }{ 2\left( M_{\rm A}+1 \right) }
                        \frac{ P_{\rm w}    }{ \rho },
\\
\label{eq:Vg}
&& V_{\rm g}{}^2 \equiv
\frac{ A }{ A' }
\left[
  \frac{ d\Phi       }{ dz }
+ \frac{ \gamma_{\rm g}-1  }{ \rho v   }{\cal H}
+ \frac{ 1                 }{ \rho }
  \frac{ dP_{\rm cr} }{ dz }
\right],
\\
\label{eq:cal H}
&& {\cal H} \equiv -n^2\Lambda - V_{\rm A}\frac{dP_{\rm cr}}{dz}.
\end{eqnarray}
%%%%%%%%%%%%%%
%
We assume $v'>0$ for the smooth outflow solution.
We will see that the role of CRs is important for this condition.
\par
First, we review the primal properties of a steady flow in a gravitational potential.
Considering the simplest case $C_{\rm w}=0$, $P_{\rm cr}=0$, $P'_{\rm cr}=0$, and ${\cal H}=0$ (i.e.,
analogous to a simple adiabatic solar wind solution) as an example, the sign of the
wind equation depends on
whether the flow is subsonic ($v<C_{\rm g}$) or supersonic ($v>C_{\rm g}$).
The subsonic flow with a negligible gravitational acceleration ($d\Phi/dz
\ll C_{\rm g}{}^2A'/A$) shows always $v'<0$. This reflects simply the adiabatic
expansion in vacuum. In contrast to this, a sufficiently large gravitational pull
($d\Phi/dz > C_{\rm g}{}^2A'/A$) corresponds to a negative numerator on the right-hand side
of Eq.~\eqref{eq:wind equation}. Since the denominator is negative in the subsonic region,
this results in $v'>0$.
The accelerated flow passes the transonic point at which the conditions of
$v^2=C_{\rm g}{}^2=V_{\rm g}{}^2$ should be satisfied so that $v'$ is finite and positive.
For a supersonic flow, the internal energy is no longer important for the fluid dynamics,
and the condition of $v'>0$ is $d\Phi/dz < C_{\rm g}{}^2A'/A$.
This means that the mean kinetic energy of particles in the fluid element $\sim m C_{\rm g}^2$
is larger than the gravitational potential energy $m\Phi$, i.e., the particles should be
gravitationally unbound.
\par
In our case, the CRs can accelerate the fluid by a
term of $vdP_{\rm cr}/dz$, and can heat the fluid
via the wave generation and dissipation ($V_{\rm A}dP_{\rm cr}/dz$).
The former makes the rate of adiabatic losses large, that is,
too much CR pressure results in $v'<0$ for a subsonic flow.
On the other hand, for a supersonic flow,
this additional pressure support helps to make $v'>0$.
The latter acts in the opposite sense. The heating acts against the radiative or adiabatic
cooling; thus a sufficient amount of CRs tends to make $v'>0$ for a subsonic flow. For a supersonic flow,
too much heating results in too high thermal pressure that dominates over the ram pressure
of the fluid.
In such a case, the condition of the smooth wind solution $v'>0$ everywhere is not satisfied due to
the existence of a pressure bump.
These effects can be seen by rewriting the wind equation.
Considering a subsonic flow as an example, from the condition of $v'>0$, we obtain
%
%%%%%%%%%%%%%%
\begin{eqnarray}
%
%~~~~~~~~~~~~~~~~ for one column ~~~~~~~~~~~~~~~~~~~~~~
%\left(\frac{\gamma_{\rm g}-1}{M_{\rm A}}-1\right)\frac{ dP_{\rm cr} }{dz}
%< \rho \frac{d\Phi}{dz} - \frac{ \gamma_{\rm g}-1 }{v}n^2\Lambda
%- \frac{ A' }{ A } \rho \left( C_{\rm g}{}^2+C_{\rm w}^2 \right).
%~~~~~~~~~~~~~~~~ for one column ~~~~~~~~~~~~~~~~~~~~~~
%
%~~~~~~~~~~~~~~~~ for two column ~~~~~~~~~~~~~~~~~~~~~~
\left(\frac{\gamma_{\rm g}-1}{M_{\rm A}}-1\right)\frac{ dP_{\rm cr} }{dz}
&<& \rho \frac{d\Phi}{dz} - \frac{ \gamma_{\rm g}-1 }{v}n^2\Lambda
\nonumber \\
&-& \frac{ A' }{ A } \rho \left( C_{\rm g}{}^2+C_{\rm w}^2 \right).
%~~~~~~~~~~~~~~~~ for two column ~~~~~~~~~~~~~~~~~~~~~~
%
\end{eqnarray}
%%%%%%%%%%%%%%
%
Thus, when $M_{\rm A}\gg \gamma_{\rm g}-1$ ($M_{\rm A}\ll \gamma_{\rm g}-1$),
the CR pressure acts as the term of $vdP_{\rm cr}/dz$ (the heating), and
$dP_{\rm cr}/dz$ has an upper (lower) limit. For a supersonic flow, the inequality
reverses. As we discussed in Sect.~\ref{sec:model}, the plasma-$\beta$ has $\beta\sim60$
at the bottom region of the outflow. Thus, the Alfv{\'e}n Mach number is about
$M_{\rm A}\sim\sqrt{\beta}\sim8\gg2/3$.
Hence, the CRs contribute to realize the steady-state outflow via the term of $vdP_{\rm cr}/dz$,
which acts as an additional pressure support for a supersonic flow.
Note that the above argument appears because we adopt the steady-solution
with $v'>0$ everywhere.
\par
Finally, we analyze the conditions of the transonic point following~\citet[][]{breitschwerdt91}.
\footnote{The original definition of $C_*$
by \citet{breitschwerdt91} includes the CR pressure by neglecting the CR diffusion.
\citet{recchia16} considered the CR diffusion and also defined the $C_*$ with the CR pressure.
In such definition, however, the expression of $C_*$ shows an apparent diverging point.
To avoid this, we treat the CR pressure separately.}
Defining $C_*{}^2\equiv C_{\rm g}{}^2+C_{\rm w}{}^2$, we rewrite the wind equation as
%
%%%%%%%%%%%%%%
\begin{eqnarray}
   \frac{ dv }{ dz }
=  \frac{ v }{ z }
   \frac{ {\cal N}(v,z) }{ (v+C_*)(v-C_*) },
\end{eqnarray}
where
\begin{eqnarray}
   {\cal N} = z
\frac{A'}{A} \left( C_*{}^2 - V_{\rm g}{}^2  \right).
\end{eqnarray}
%%%%%%%%%%%%%%
%
At the transonic point $z=z_{\rm tr}$, the conditions $v^2=C_*{}^2$ and
$C_*{}^2=V_{\rm g}{}^2$ should be satisfied simultaneously so that
the velocity gradient is finite and positive.
We obtain the gradient at $z_{\rm tr}$ by the linearization of the
equation around the transonic point as
%
%%%%%%%%%%%%%%
\begin{eqnarray}
&& \frac{ dv }{ dz }
\approx
\frac{ v }{ 2z_{\rm tr} }
\frac{ 1 }{ v - C_* }
\left[ (z-z_{\rm tr}){\cal N}_z + (v-v_{\rm tr}){\cal N}_v \right],
\end{eqnarray}
%%%%%%%%%%%%%%
%
where ${\cal N}_z = ( \partial {\cal N} / \partial z )_v$
and   ${\cal N}_v = ( \partial {\cal N} / \partial v )_z$, respectively.
Then, introducing the transformations of
%
%%%%%%%%%%%%%%
\begin{eqnarray}
\eta = \frac{ v - C_* }{ C_* }~{\rm and}~\zeta = \frac{ z-z_{\rm tr} }{ z_{\rm tr} },
\nonumber
\end{eqnarray}
%%%%%%%%%%%%%%
%
we obtain
%
%%%%%%%%%%%%%%
\begin{eqnarray}
&& \eta\frac{ d\eta }{ d\zeta } = \alpha\zeta + \beta\eta,
\\
&& \alpha = \frac{ z_{\rm tr} }{ 2C_* }{\cal N}_z,
\\
&& \beta  = \frac{ 1          }{ 2C_* }{\cal N}_v.
\end{eqnarray}
%%%%%%%%%%%%%%
%
Substituting $\eta=w\zeta$, we find the solution of this differential equation as
%
%%%%%%%%%%%%%%
\begin{eqnarray}
\label{eq:grad v transonic}
&& \left( \frac{ dv }{ dz } \right)_{\rm tr}
= \frac{C_*}{z_{\rm tr}}w,
\\
&& w
= \frac{ \beta \pm \sqrt{ \beta^2 + 4\alpha } }{2}.
\label{eq:grad v tr}
\end{eqnarray}
%%%%%%%%%%%%%%
%
Since we consider the outflow solution ($dv/dz > 0$), we take
the positive sign in front of the square root of $w$.
We show the expressions of ${\cal N}_z$ and ${\cal N}_v$ in the appendix.
\par
To find the wind solution, we set the values ($v$, $\rho$, $P_{\rm g}$, $P_{\rm cr}$, $dP_{\rm cr}/dz$,
$P_{\rm w}$, and $B$) at the transonic point as a boundary condition. The location of the transonic point
is given by the condition of $v^2=C_*{}^2=V_{\rm g}{}^2$ because $V_{\rm g}{}^2$ contains the term of
$d\Phi/dz$, which is a given function of $z$. The velocity gradient at the transonic point is calculated
by using Eq.~\eqref{eq:grad v tr}, and then the gradients of other values ($dP_{\rm g}/dz$,
$d^2P_{\rm cr}/d^2z$, and $dP_{\rm w}/dz $) are obtained. Integrating these differential equations
from the transonic point toward the top and bottom boundaries, we can find a smooth solution
with $v'>0$ at arbitrary $z$. Note that this method ensures the flow passing through the transonic
point; however, it is not guaranteed whether the solution continues at the both boundaries with a positive
and finite $v'$. For example, some boundary conditions lead to a flow showing $v^2\rightarrow C_*{}^2\neq
V_{\rm g}{}^2$ at the subsonic region between the transonic point and bottom boundary that results in a divergence
of the velocity gradient. Such behavior results from the effects of the radiative cooling, and the topology of the flow
is different from the case of no radiative cooling~\citep[e.g.,][]{breitschwerdt91,recchia16}. We can efficiently
exclude such `failed' solutions by starting from
the transonic point.

\section{Results}
\label{sec:results}
To derive the outflow solutions, we set boundary conditions at the transonic point $z=z_{\rm tr}$.
We fix two parameters as $B_{\rm tr}=1~{\rm \mu G}$ and $P_{\rm w,tr}=1.56\times10^{-8}~{\rm eV~cm^{-3}}$
(i.e., $\kappa_{\rm tr}=5.3\times10^{28}~{\rm cm^2~s^{-1}}$).
The subscript `tr' indicates the values at the transonic point.
We have chosen logarithmically spaced $N$ different values of other parameters for
$n_{\rm tr}, T_{\rm tr}, P_{\rm cr,tr}$, and $H_{\rm cr,tr}$, that can be expressed
by the following formula:
%
%%%%%%%%%%%%%
\begin{eqnarray}
f = \log f_{\rm m} + \frac{i-1}{N}\log\frac{f_{\rm M}}{f_{\rm m}}~~~(i=1\mathchar`-N),
\label{eq:parameter}
\end{eqnarray}
%%%%%%%%%%%%%
%
where $f$ symbolically indicates $n_{\rm tr}$, $T_{\rm tr}$, $P_{\rm cr,tr}$, and $H_{\rm cr,tr}$.
The values of $f_{\rm m}$, $f_{\rm M}$, and $N$ are summarized in Table~\ref{tab:parameter}.
%
%%%%%%%%%%%%
\begin{table}
\tabletypesize{\scriptsize}
\centering
\caption{
Parameter range of the boundary conditions, where
$n_{\rm tr}$, $T_{\rm tr}$, $P_{\rm cr,tr}$, and $H_{\rm cr,tr}$ are
the number density, temperature, pressure of CRs, and scale height of CRs
($H_{\rm cr,tr}\equiv P_{\rm cr,tr} \big|dP_{\rm cr,tr}/dz\big|^{-1}$)
at the transonic point. Each parameter range from $f_{\rm m}$ to $f_{\rm M}$
divided by $N$ in logarithmically space (see, equation \ref{eq:parameter}). 
}
\label{tab:parameter}
\begin{tabular}{ccccc}
\hline
            & $n_{\rm tr}$ & $T_{\rm tr}$ & $P_{\rm cr,tr}$ & $H_{\rm cr,tr}$ \\
\hline
$f_{\rm m}$ & $10^{-3}$~cm$^{-3}$  & $10^6$~K & $0.1$~eV~cm$^{-3}$ & $ 10$~kpc  \\
$f_{\rm M}$ & $10^{-2}$~cm$^{-3}$  & $10^7$~K & $ 10$~eV~cm$^{-3}$ & $300$~kpc  \\
$N$         & $16$                 & $16$     & $16$               & $32$       \\
\hline
\end{tabular}
\end{table}
%%%%%%%%%%%%
%
Note that $dP_{\rm cr,tr}/dz=-P_{\rm cr,tr}/H_{\rm cr,tr}$.
Then, we integrate
the Eqs.~\eqref{eq:basic equation of motion}-\eqref{eq:basic wave energy}
from the transonic point toward the bottom boundary $z_{\rm bt}=2~{\rm kpc}$
and toward the top boundary $z_{\rm tp}=350~{\rm kpc}$ by the fourth Runge-Kutta method,
respectively. For those boundary conditions, successful outflow solutions we consider satisfy
that $v'>0$, $P_{\rm cr}'<0$, and $T>10^4~{\rm K}$ at $z_{\rm bt}<z<z_{\rm tp}$, and $v<C_*$
at $z_{\rm bt}<z<z_{\rm tr}$, and $v>C_*$ at $z_{\rm tr}<z<z_{\rm tp}$,
where the subscripts `bt' and `tp' denote the values at the bottom boundary
and top boundary, respectively.
We apply this procedure for 10 horizontal positions, $R=(1; 2; 3; 4; 5; 6; 7; 8; 9; 10 )$~kpc,
and find $327,254$ solutions in total as a result.
\par
%
%%%%%%%%%%%%%%
\begin{figure}[htbp]
\centering
\includegraphics[scale=0.7]{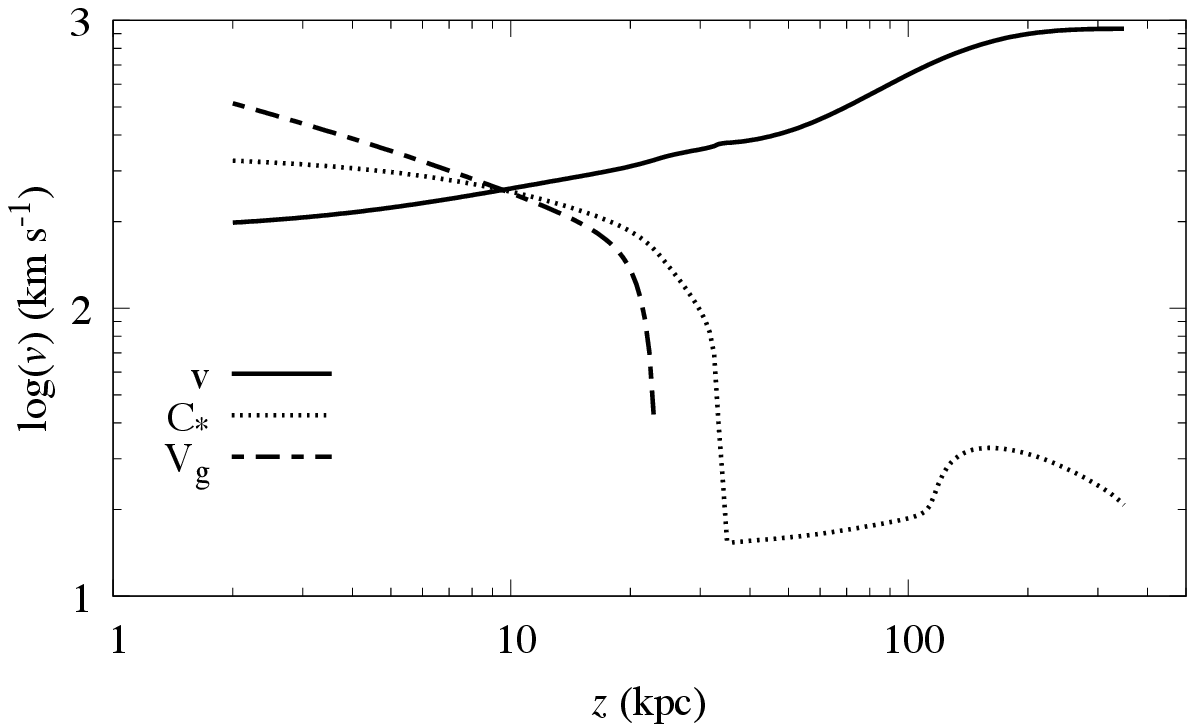}
\includegraphics[scale=0.7]{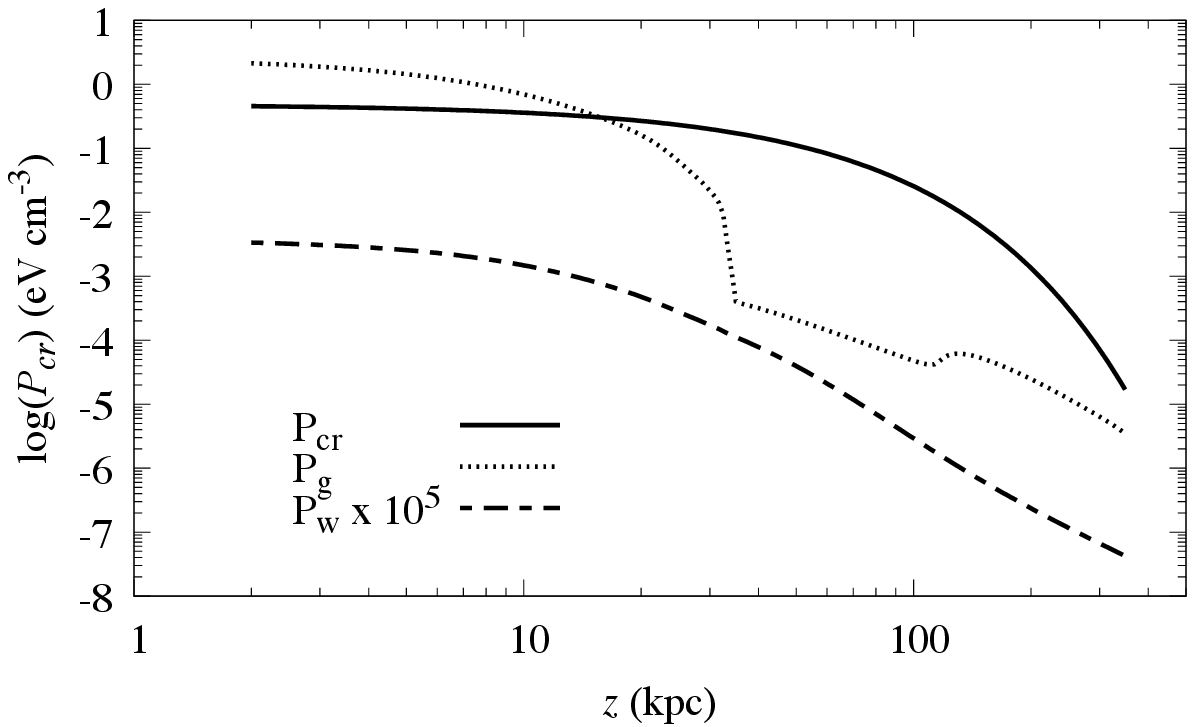}
\caption{
Solution at $R=8$~kpc for
$n_{\rm tr}    \simeq 2.74\times10^{-3}~{\rm cm^{-3}}$,
$T_{\rm tr}    \simeq 2.74\times10^{6}~{\rm K}$,
$P_{\rm cr,tr} \simeq 0.316~{\rm eV~cm^{-3}}$, and
$H_{\rm cr,tr} \simeq 32.2~{\rm kpc}$.
(top panel): The solid line shows the velocity $v$, the dots show $C_*$, and
the dashed line shows $V_{\rm g}$, respectively.
Note that $C_*\simeq C_{\rm g}=\sqrt{\gamma_{\rm g}P_{\rm g}/\rho}$.
(bottom panel): The solid line shows the CR pressure $P_{\rm cr}$,
the dots show the thermal pressure $P_{\rm g}$, and the dashed shows
times $10^5$ the pressure of the Alfv{\'e}n waves $P_{\rm w}$.
}
\label{fig:single velocity}
\end{figure}
%%%%%%%%%%%%%%
%
We discuss the nature of the outflow using the solution
at $R=8$~kpc for
$n_{\rm tr}    \simeq 2.74\times10^{-3}~{\rm cm^{-3}}$,
$T_{\rm tr}    \simeq 2.74\times10^{6}~{\rm K}$,
$P_{\rm cr,tr} \simeq 0.316~{\rm eV~cm^{-3}}$, and
$H_{\rm cr,tr} \simeq 32.2~{\rm kpc}$ as a representative result.
Note that $P_{\rm g,tr}=n_{\rm tr}kT_{\rm tr}
\simeq0.614~{\rm eV~cm^{-3}}$.
The values at the bottom boundary of this solution are
$n_{\rm bt}    \simeq 5.23\times10^{-3}~{\rm cm^{-3}}$,
$T_{\rm bt}    \simeq 4.70\times10^{6}~{\rm K}$,
$P_{\rm cr,bt} \simeq 0.456~{\rm eV~cm^{-3}}$,
$H_{\rm cr,bt} \simeq 33.9~{\rm kpc}$, and
$B_{\rm bt}    \simeq 1.44~{\rm \mu G}$, respectively.
The top panel of Fig.~\ref{fig:single velocity}
shows the velocities of $v$ (solid line), $C_*$ (dots), and $V_{\rm g}$ (dashed line).
Note that $C_*\simeq C_{\rm g}=\sqrt{\gamma_{\rm g}P_{\rm g}/\rho}$.
The transonic point is located at $z_{\rm tr}\simeq9.54$~kpc.
$V_{\rm g}{}^2$ becomes
negative at $z\sim20$~kpc because in Equation~\eqref{eq:Vg},
the term of $vdP_{\rm cr}/dz$
dominates over the gravitational acceleration (i.e., $v'\approx P_{\rm cr}'/\rho v$).
$C_*$ rapidly decreases from $z\sim20$~kpc due to the radiative cooling.
The minimum temperature is about $1.043\times10^4$~K at which $\beta\simeq1.070$.
Then, the cooling is balanced by the heating due to the wave dissipation.
Since the cooling rate $n^2\Lambda$ decreases with increasing the velocity and area
cross section of the flux tube ($\rho \propto 1/ v A$), the outflow is slightly heated
due to the wave dissipation.
The outflow begins to be adiabatically cooled again from $z\sim100$~kpc because
the CR pressure drops exponentially. This exponential decay of $P_{\rm cr}$
results from the effects of the CR diffusion.
Note that, if we consider a case of no diffusion of CRs ($\kappa=0$), the CR transport
equation~\eqref{eq:basic cosmic-ray transport} could be rewritten as
$P_{\rm cr}\propto[(v+V_{\rm A})A]^{-\gamma_c}$ \citep[e.g.,][]{breitschwerdt91}.
The bottom panel of Fig.~\ref{fig:single velocity} shows the pressures of
$P_{\rm cr}$ (solid line), $P_{\rm g}$ (dots), and $P_{\rm w}\times10^{5}$ (dashed line).
The CR pressure is well represented by $\exp(-z/H_{\rm cr,bt})$ for this solution.
\par
%
%%%%%%%%%%%%%%
\begin{figure}[htbp]
\centering
\includegraphics[scale=0.7]{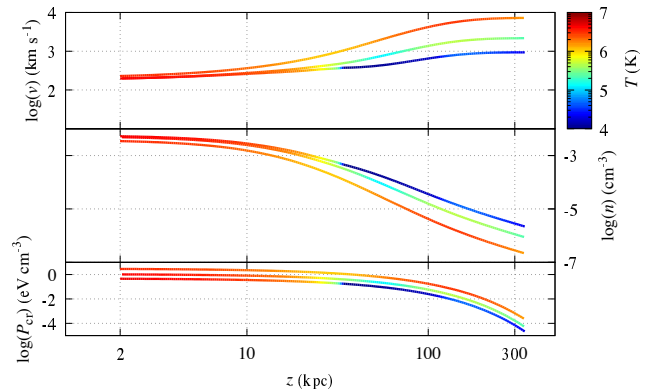}
\caption{
Solutions at $R=8$~kpc for
$P_{\rm cr,tr} \simeq 0.316~{\rm eV~cm^{-3}}$ (blue),
$P_{\rm cr,tr} \simeq 0.750~{\rm eV~cm^{-3}}$ (green), and
$P_{\rm cr,tr} \simeq 2.37 ~{\rm eV~cm^{-3}}$ (red).
The other parameters are the same as in Fig.~\ref{fig:single velocity}.
The top panel shows the velocity, middle panel shows the number density,
and bottom panel shows the CR pressure. The color indicates the temperature.
}
\label{fig:three solutions}
\end{figure}
\begin{figure}[htbp]
\centering
\includegraphics[scale=0.7]{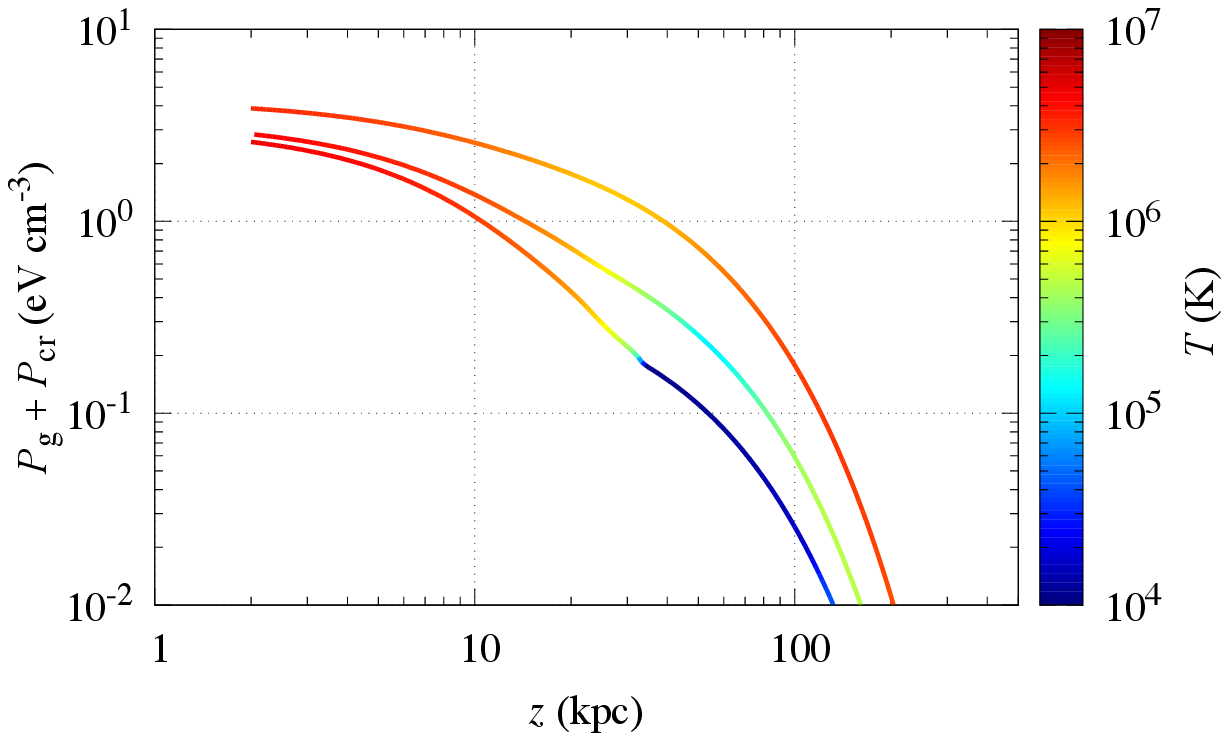}
\includegraphics[scale=0.7]{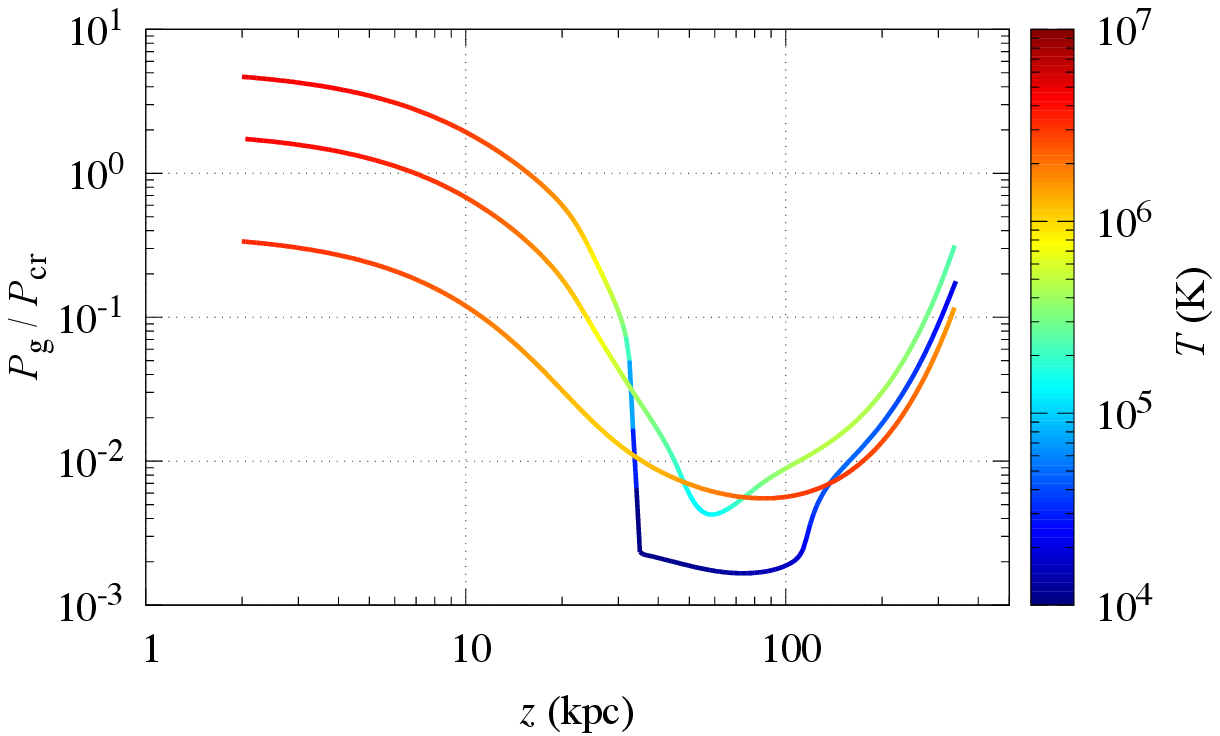}
\caption{
Total pressure $P_{\rm g}+P_{\rm cr}$ (top panel)
and ratio of the thermal pressure to CR pressure (bottom panel)
for the solutions that are the same as those in Fig.~\ref{fig:three solutions}.
The color represents the temperature.
}
\label{fig:ratio of pressure}
\end{figure}

%%%%%%%%%%%%%%
%
We consider the effects of the amount of CRs. Figure~\ref{fig:three solutions}
shows three solutions at $R=8$~kpc for
$P_{\rm cr,tr} \simeq 0.316~{\rm eV~cm^{-3}}$ (blue),
$P_{\rm cr,tr} \simeq 0.750~{\rm eV~cm^{-3}}$ (green), and
$P_{\rm cr,tr} \simeq 2.37 ~{\rm eV~cm^{-3}}$ (red).
The other parameters are the same as in Fig.~\ref{fig:single velocity}.
Table~\ref{tab:bottom three} summarizes the values at the bottom boundary
for these three solutions.
%
%%%%%%%%%%%%
\begin{table}
\tabletypesize{\scriptsize}
\centering
\caption{
The values at the bottom boundary $z_{\rm bt}$
for the solutions for $P_{\rm cr,tr}=0.316$,
$0.750$, and $2.37~{\rm eV~cm^{-3}}$.
The other boundary conditions are the same as the solution shown in
Fig.~\ref{fig:single velocity}.
}
\label{tab:bottom three}
\begin{tabular}{cccccc}
\hline
  $P_{\rm cr,tr}$
& $n_{\rm bt}$    
& $T_{\rm bt}$    
& $P_{\rm g,bt}$  
& $P_{\rm cr,bt}$ \\
%& $\rho v$ \\
  ${\rm eV cm^{-3}}$
& $10^{-3}~{\rm cm^{-3}}$
& $10^6$~K   
& ${\rm eV~cm^{-3}}$
& ${\rm eV~cm^{-3}}$ \\
%& ${\rm eV~cm^{-3}}$ \\
\hline
0.316     & 5.23  & 4.70  & 2.12  & 0.456 \\ %& 6.66E-8  \\
0.750     & 4.85  & 4.30  & 1.80  & 1.04  \\ %& 6.14E-8  \\
2.37      & 3.46  & 3.25  & 0.969 & 2.90  \\ %& 5.06E-8  \\
\hline 
\end{tabular}
\end{table}
%%%%%%%%%%%%
%
As shown in Table~\ref{tab:bottom three},
the thermal pressure $P_{\rm g,bt}$ is anticorrelated to the CR pressure
$P_{\rm cr,bt}$. Since the acceleration by large pressures
overwhelming the gravitational acceleration at the subsonic region
makes $v'<0$ (see Section~\ref{sec:wind equation}), the total pressure
may be regulated to a certain degree to satisfy the given boundary conditions
at the transonic point. Note that the thermal pressures at the transonic points
are the same as each of the three solutions (i.e. the total pressures are different);
therefore, the regulation of the total pressure at the bottom boundary is not so obvious.
Figure~\ref{fig:ratio of pressure}
shows the total pressure $P_{\rm g}+P_{\rm cr}$ (top panel)
and ratio of the pressures $P_{\rm g}/P_{\rm cr}$ (bottom panel)
for these three solutions. 
%
%%%%%%%%%%%%%%
%%%%%%%%%%%%%%
%
%
%%%%%%%%%%%%%%
\begin{figure}
\centering
\includegraphics[scale=0.7]{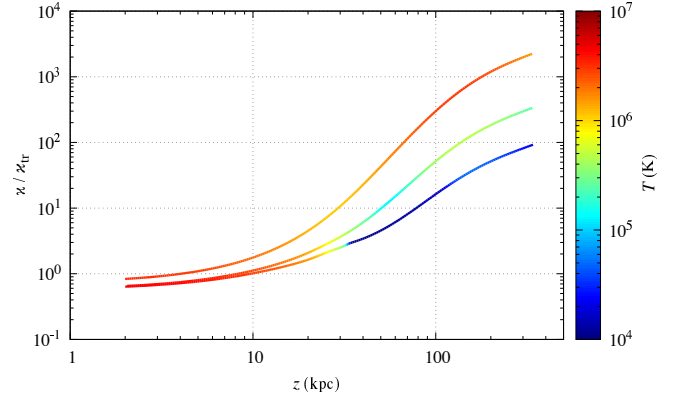}
\caption{
Spatial profile of the diffusion coefficient $\kappa(z)/\kappa_{\rm tr}$
at $R=8$~kpc for $P_{\rm cr,tr} \simeq 0.316~{\rm eV~cm^{-3}}$ (blue),
$P_{\rm cr,tr} \simeq 0.750~{\rm eV~cm^{-3}}$ (green), and
$P_{\rm cr,tr} \simeq 2.37 ~{\rm eV~cm^{-3}}$ (red).
The other parameters are the same as in Figure~\ref{fig:single velocity}.
The color indicates the temperature.
}
\label{fig:kappa}
\end{figure}
%%%%%%%%%%%%%%
%
In the case of the relatively small CR pressure (blue line), a relatively
large thermal pressure $P_{\rm g}=nkT$ is required. Since the temperature
should be around the virial temperature, this requirement results in
a relatively large density. Then, the outflow suffers the radiative
cooling at a rate of $n^2\Lambda$, and the CR pressure becomes the dominant
component to accelerate the outflow. Contrary to this, when the CR
pressure is relatively large (red line), the required thermal pressure
becomes low, which results in a relatively small density.
Since the cooling rate depends on $n^2$ and the heating rate depends on
$P_{\rm cr}$, the temperature (and thermal pressure)
remains to be high. Thus, a smaller (larger) CR pressure results
in a slower and colder (faster and hotter) outflow.
Note that the amount of CRs at a local point is also determined by the diffusion
coefficient. Figure~\ref{fig:kappa} shows the spatial profile of the coefficient
$\kappa(z)/\kappa_{\rm tr}$. The hotter wind results from a larger $\kappa$ because
the CRs can reach at a higher $z$.
\par
%
%%%%%%%%%%%%%%
\begin{figure}[htbp]
\centering
\includegraphics[scale=0.7]{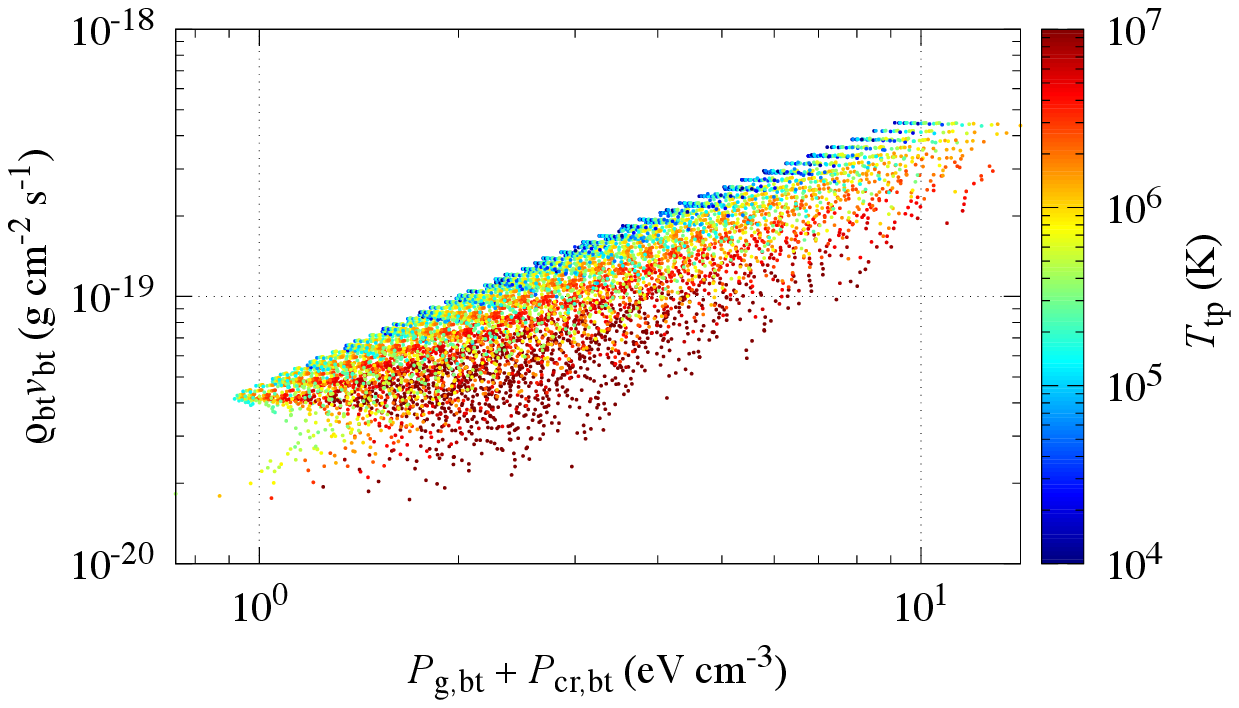}
\includegraphics[scale=0.7]{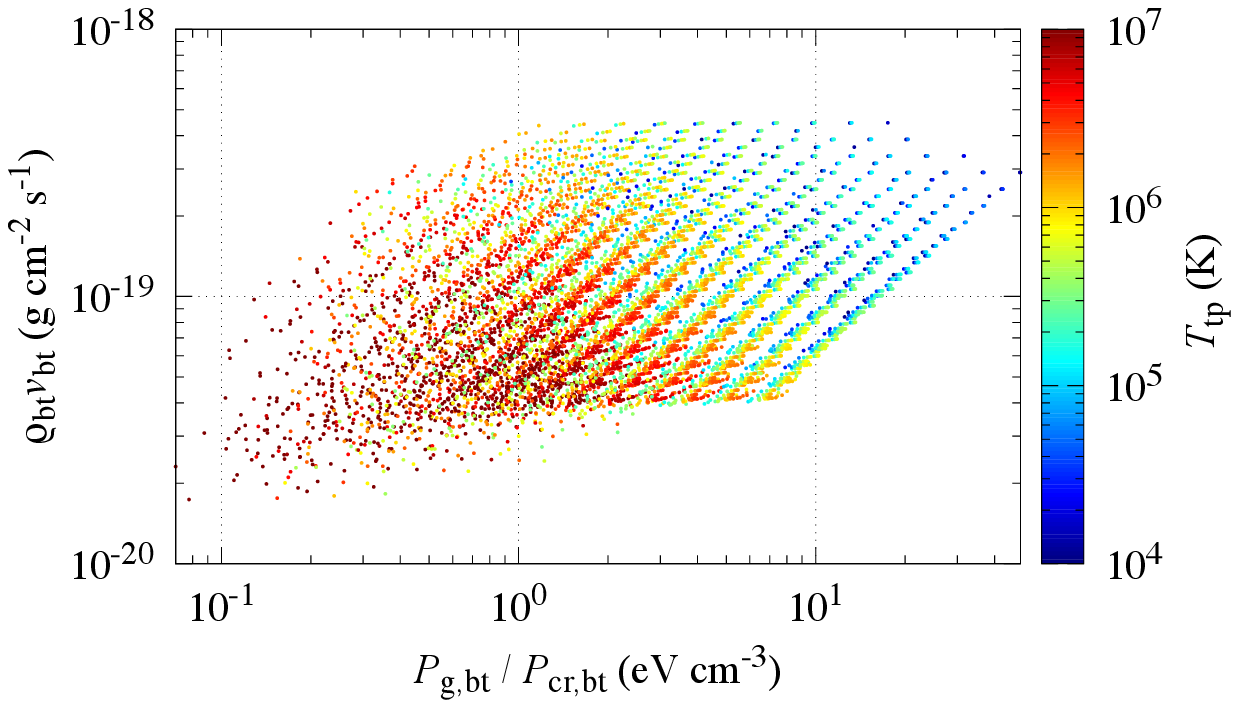}
\caption{
(top panel): The relation between
the total pressure $P_{\rm g,bt}+P_{\rm cr,bt}$
and mass flux $\rho_{\rm bt}v_{\rm bt}$.
(bottom panel): The relation between
the ratio of pressures $P_{\rm g,bt}/P_{\rm cr,bt}$
and mass flux $\rho_{\rm bt}v_{\rm bt}$.
The color shows the temperature $T_{\rm tp}$ at the top boundary
($z=z_{\rm tp}$).
}
\label{fig:P rhov Ttp}
\end{figure}
%%%%%%%%%%%%%%
%
%
%%%%%%%%%%%%%%
\begin{figure}[htbp]
\centering
\includegraphics[scale=0.7]{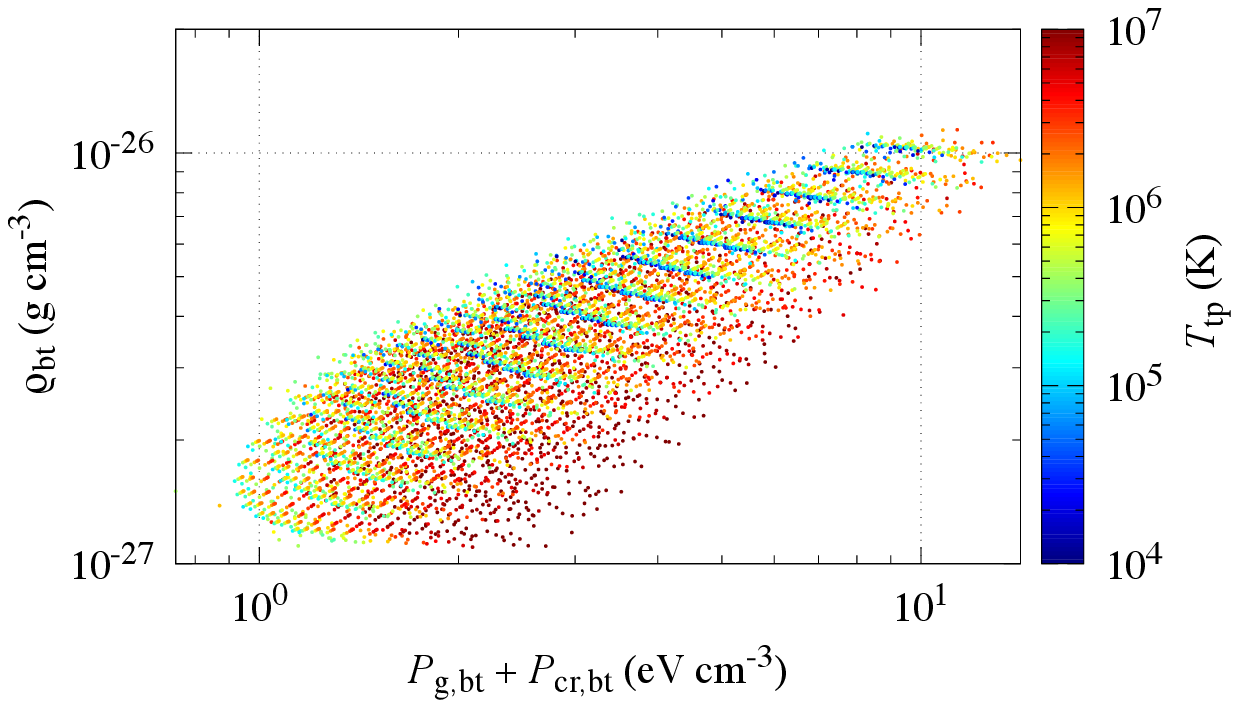}
\includegraphics[scale=0.7]{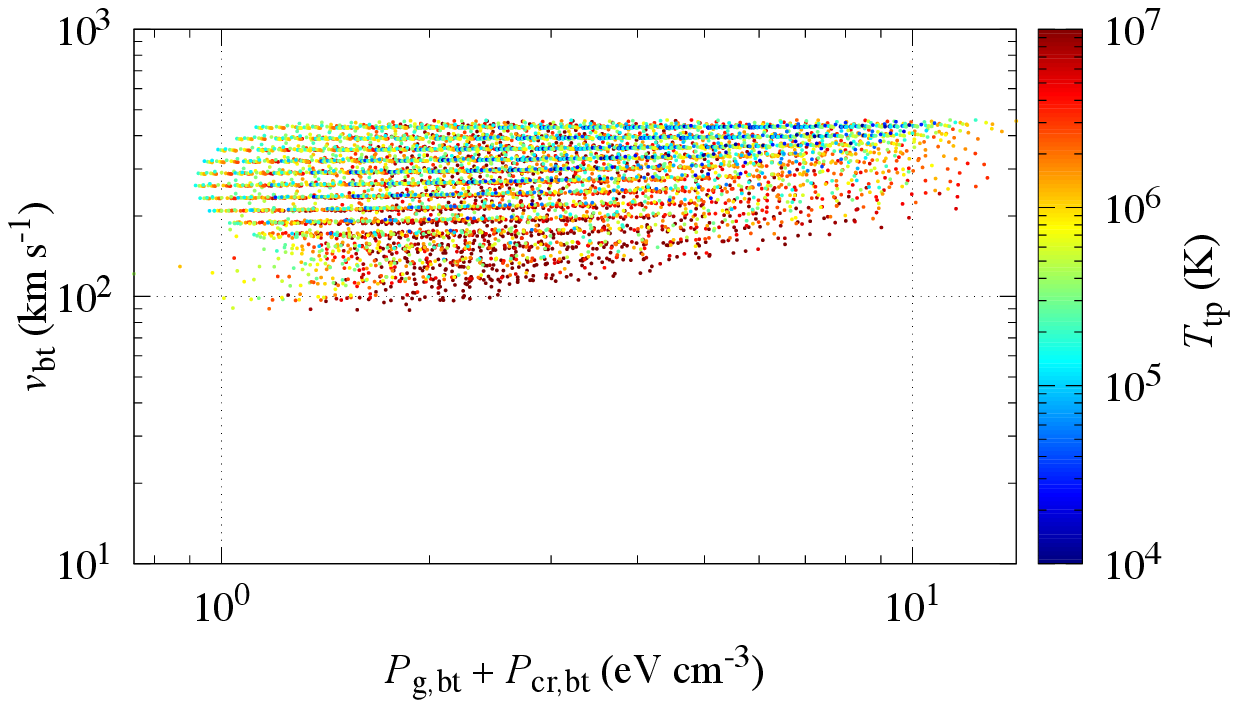}
\caption{
The relations between
the total pressure $P_{\rm g,bt}+P_{\rm cr,bt}$
and density $\rho_{\rm bt}$ (top panel), and
between the total pressure and
velocity $v_{\rm bt}$ (bottom panel).
The color shows the temperature $T_{\rm tp}$ at the top boundary
($z=z_{\rm tp}$).
}
\label{fig:P rho Ttp}
\end{figure}
%%%%%%%%%%%%%%
%
Here, we confirm the above discussion from the overall trend of the solutions.
Figure~\ref{fig:P rhov Ttp} shows the values at the bottom and top
boundaries for the derived solutions at $R=8$~kpc.
The top panel shows the relation between
$P_{\rm g,bt}+P_{\rm cr,bt}$, $\rho_{\rm bt}v_{\rm bt}$,
and $T_{\rm tp}$.
The total pressure range is $1\mathchar`-10~{\rm eV~cm^{-3}}$.
The mass flux shows a roughly linear dependence on
the total pressure. As shown in Figure~\ref{fig:P rho Ttp},
this mainly results from the dependence of $\rho_{\rm bt}$,
while $v_{\rm bt}$ does not depend on the total pressure
(a scattering of $v_{\rm bt}$ is about a factor of 2).
Since the temperature should be around the virial temperature (indeed our solutions
show $T_{\rm bt}\sim10^6\mathchar`-10^7$~K$\sim T_{\rm vir}$),
this dependence simply reflects the equation of state $P_{\rm g}=n kT$.
Then, a faster outflow at the subsonic region can reach a larger height,
at which the gravitational acceleration is negligible compared with
the rate of the adiabatic cooling, so it tends to make $v'<0$.
Thus, the larger pressure results in a more massive outflow with fixed velocity
rather than accelerating the outflow with fixed mass.
The bottom panel of Fig.~\ref{fig:P rhov Ttp} shows the relation between
$P_{\rm g,bt}/P_{\rm cr,bt}$, $\rho_{\rm bt}v_{\rm bt}$,
and $T_{\rm tp}$.
As we discussed above,
a relatively small (large) CR pressure results in a cold (hot) outflow.
Thus, the mass flux is determined by
the total pressure, while the properties of the outflow (especially the temperature)
are controlled by the amount of CRs.
\par
We discuss the fate of the outflow in the three cases summarized in
Table~\ref{tab:bottom three}.
%
%%%%%%%%%%%%%%
\begin{figure}[htbp]
\centering
\includegraphics[scale=0.7]{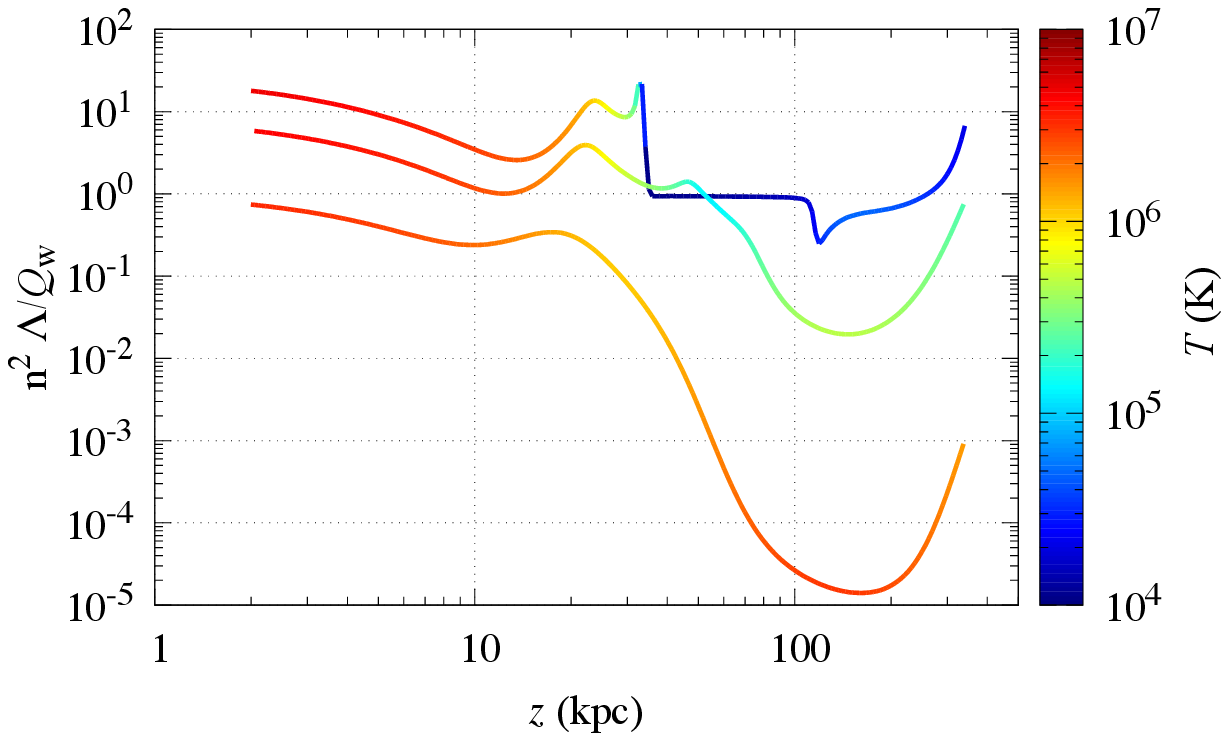}
\includegraphics[scale=0.7]{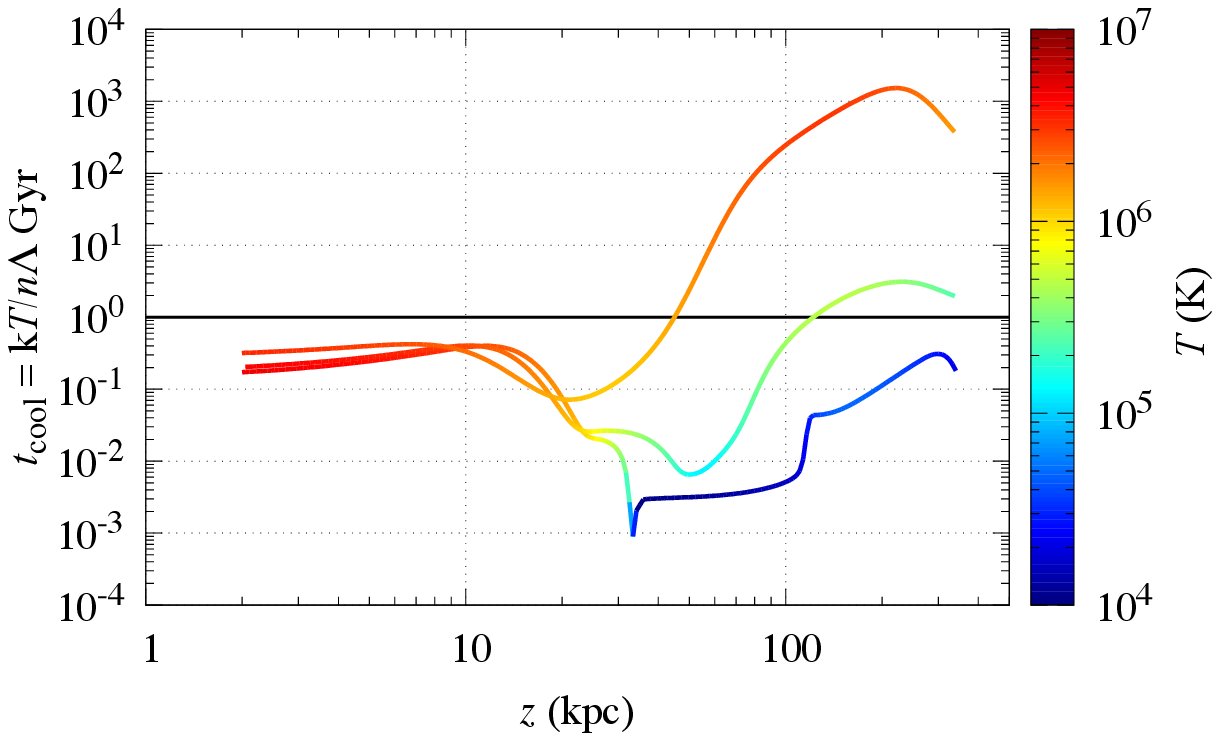}
\caption{
Ratio of the cooling rate to heating rate $n^2\Lambda/Q_{\rm w}$ (top panel)
and cooling time $t_{\rm cool}=kT/n\Lambda$ (bottom panel)
for the solutions that are the same as those in Figure~\ref{fig:three solutions}.
The color represents the temperature.
}
\label{fig:cooling}
\end{figure}
%%%%%%%%%%%%%%
%
Figure~\ref{fig:cooling} shows the ratio of the cooling rate to the heating rate
$n^2\Lambda/Q_{\rm w}$ (top panel) and the cooling time $t_{\rm cool}
=kT/n\Lambda$ (bottom panel). Since the heating due to the wave dissipation ceases
with the exponential decrement in the CR pressure, the gas eventually cools. 
Considering the solution with $P_{\rm cr,tr}\simeq0.316~{\rm eV~cm^{-3}}$
(blue) as an example, the cooling time is shorter than $1$~Gyr.
Such cooled gas eventually decelerates due to the gravitational pull, that is,
we cannot regard that the solution continues to infinity. It may be difficult to
think that such solution represents a stable and steady-state solution, rather the solution may be
related to phenomena for which the gas falls back to the Galactic disk.
In the following, we denote the solutions with $t_{\rm cool}<1$~Gyr and
$n^2\Lambda/Q_{\rm w}>1$ at the top boundary as `fall back' solutions.
Note that, since the number density becomes smaller than $10^{-5}~{\rm cm^{-3}}$ at $z\ga100$~kpc,
the radiative heating rate can dominate over the cooling rate at $T\ga2\times10^4$~K (see
section~\ref{sec:radiative cooling}). The wind could be isothermal with $T\sim2\times10^4$, and
the isothermal temperature increases (decreases) with decreasing (increasing) the number density.
Thus, if the `fall back' phenomena are related to a condensation of gas, the radiative heating
may not affect the expectation of the `fall back' phenomena. Condensation and precipitation
are indeed likely processes of a thermally unstable gas.
\par
Finally, we estimate the total mass carried by the outflow per unit time as
%
%%%%%%%%%%%%%%%
\begin{eqnarray}
&& \Delta\dot{M}_i\equiv (\rho v)_i \times 2\pi R_i\Delta R_i \times 2, \\
&& \dot{M}_{\rm tot} = \sum_{i=1}^{10}\Delta\dot{M}_i \\
&& R_i = i~{\rm kpc},~~~\Delta R_i = 1~{\rm kpc},
\end{eqnarray}
%%%%%%%%%%%%%%%
%
where $i$ denotes the horizontal location as $R_i=i~{\rm kpc}$.
The carried mass per unit time at a radius of $R=R_i$ is
$\Delta\dot{M}_i$, which is estimated from the mass flux $\rho v$ integrated
along the axial symmetric ring at the $R=R_i$ with a width of $\Delta R_i=1~{\rm kpc}$.
The factor of 2 indicates the two directions of the outflow, $+z$ and $-z$.
The total carried mass per unit time is $\dot{M}_{\rm tot}$.
Since we suppose the axial symmetry, this estimation may give an upper limit
of the mass carried by the outflow.
Note that $\Delta\dot{M}_i$ and $\dot{M}_{\rm tot}$ were defined
in a way similar to previous studies~\citep[e.g.,][]{breitschwerdt91}.
Since we have many possible sets of solutions, we derive a statistical average.
%
%%%%%%%%%%%%%%
\begin{figure}[htbp]
\centering
\includegraphics[scale=0.7]{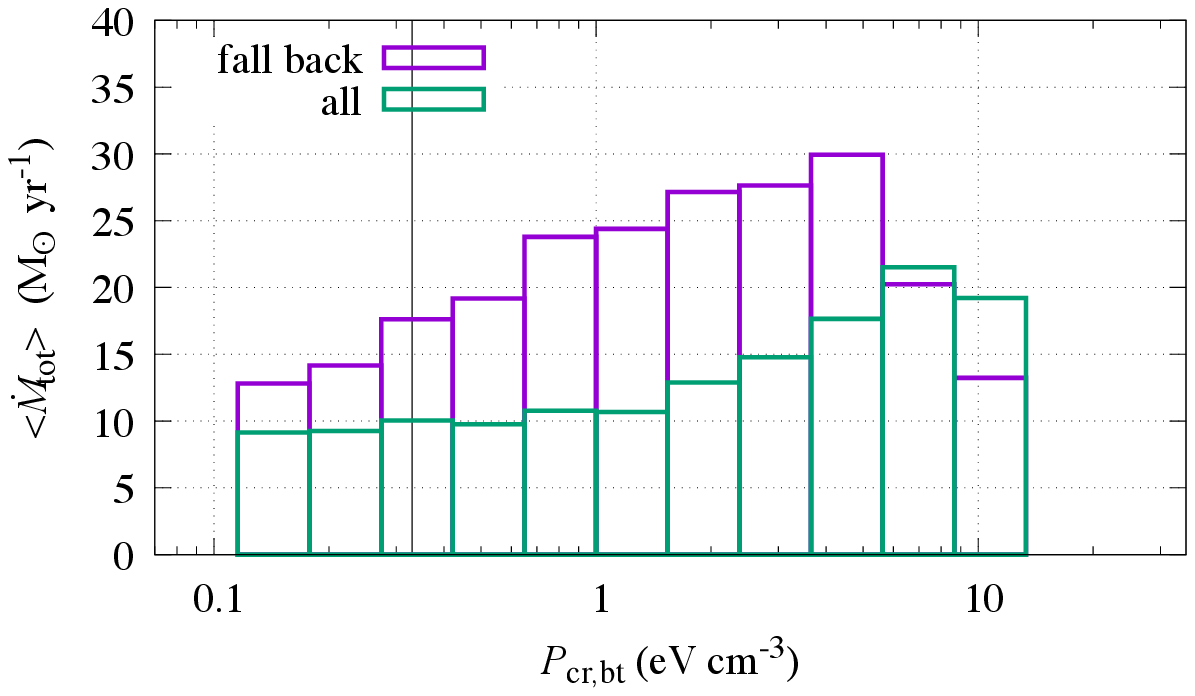}
\includegraphics[scale=0.7]{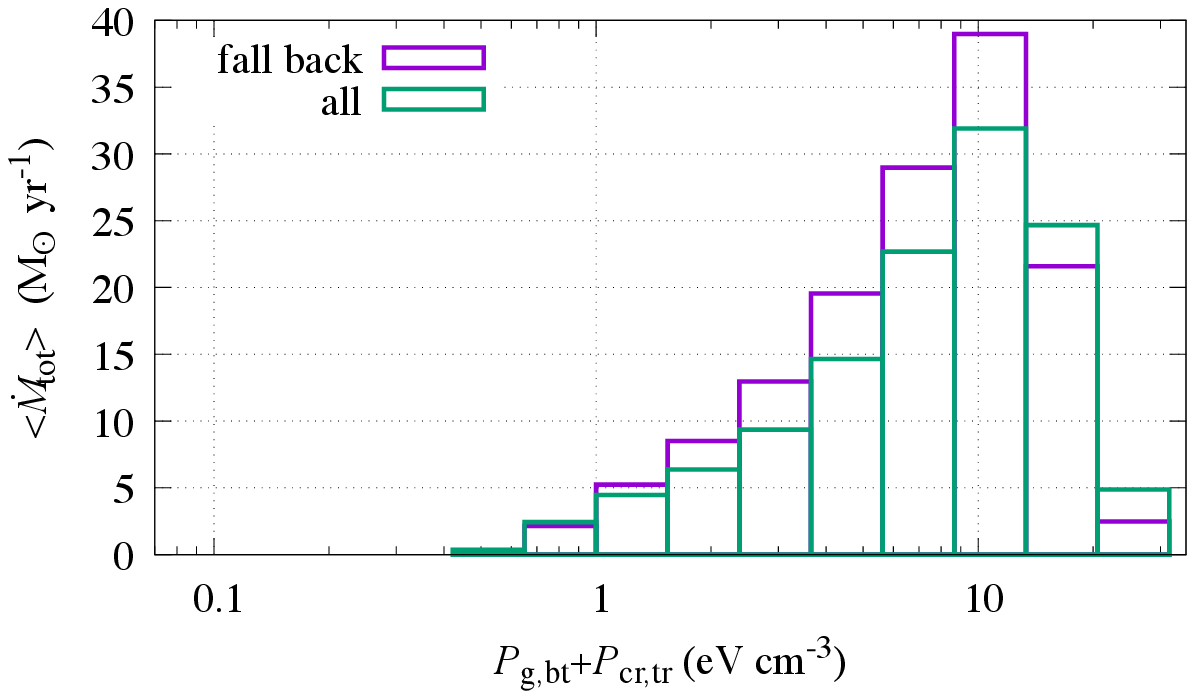}
\caption{
The statistically averaged total mass carried by the outflow per unit time
as a functions of the CR pressure (top panel) and total pressure (bottom panel)
at the bottom boundary.
The green boxes show the average for all the solutions.
The purple boxes show the average for the `fall back' solutions.
The vertical thin line indicates $P_{\rm cr,bt}=0.33~{\rm eV~cm^{-3}}$, which
corresponds to the observed CR energy density around the solar system.
}
\label{fig:total mass}
\end{figure}
%%%%%%%%%%%%%%
%
Figure~\ref{fig:total mass} shows the average of $\dot{M}_{\rm tot}$ as a function
of the CR pressure at the bottom boundary. The green boxes indicate the average
for all the solutions.
As we discussed above, the mass transfer rate does not depends sensitively on
the CR pressure; it depends on the thermal pressure (see also Fig.~\ref{fig:P rhov Ttp}).
The $13.7$~\% solutions are the `fall back' solutions. The `fall back' outflow
tends to be massive because an efficient radiative cooling is required.
%
%%%%%%%%%%%%%%
\begin{figure}[htbp]
\centering
\includegraphics[scale=0.7]{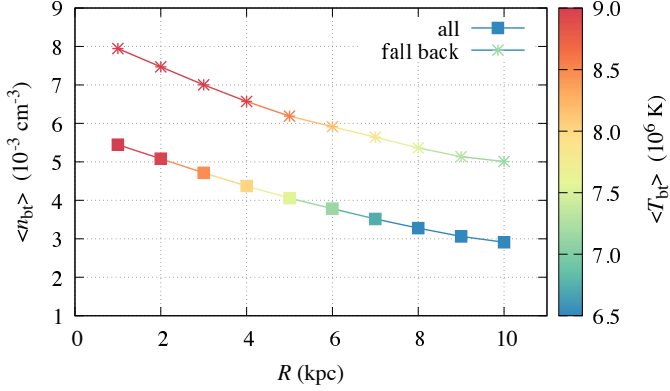}
\caption{
The statistically averaged number density at the bottom boundary
$\langle n_{\rm bt} \rangle$
as a function of the horizontal position $R$
for a CR pressure of $P_{\rm cr,bt}=0.274\mathchar`-0.422~{\rm eV~cm^{-3}}$.
The squares and asterisks indicate the averages of all solutions
and `fall back' solutions, respectively. 
The color represents the averaged temperature at the bottom boundary,
$\langle T_{\rm bt} \rangle$.
}
\label{fig:radial density}
\end{figure}
%%%%%%%%%%%%%%
%
Figure~\ref{fig:radial density} shows the averaged number density and temperature
at the bottom boundary for the CR pressure of $P_{\rm cr,bt}=0.274\mathchar`-0.422~{\rm eV~cm^{-3}}$.
The average of number density for the `fall back' solutions is roughly 2 times thath
for all the solutions. This is also indicated by the mass transfer rate. The temperature is around
the virial temperature in both cases.
The boundary conditions are consistent with
the X-ray observations~\citep[e.g.,][]{nakashima18} and numerical simulation~\citep{girichidis18};
therefore the `fall back' solutions are expected to be realized.

\section{Summary and Discussion}
\label{sec:discussion}
We have solved the steady state-Galactic wind including the effects of
the radiative cooling, CR diffusion, and heating due to the dissipation of Alfv{\'e}n
waves that are excited by the CRs. We have found 327,254 solutions in total.
For the dynamics of the thermal gas, the CR pressure
acts as the term of $vdP_{\rm cr}/dz$ or heating depending on the Alfv{\'e}n Mach number.
The former is dominant for $M_{\rm A}\gg\gamma_{\rm g}-1$, while the latter
is dominant for $M_{\rm A}\ll\gamma_{\rm g}-1$. The mass flux of the outflow
mainly depends on the thermal pressure. The CR pressure determines the
cooling and heating balance. When the CR pressure is smaller than
the thermal pressure, the outflow tends to have a lower temperature
at the vertical height of $z\sim100$~kpc. A fraction of the solutions
have a short cooling time of $<1$~Gyr with the cooling rate larger than
the heating rate at the top boundary $z=z_{\rm tp}=350$~kpc.
For these solutions, $P_{\rm g,bt}/P_{\rm cr,bt}\ga1$ is required.
We have regarded that such an outflow finally falls back to the Galactic disk.
The total mass carried by the outflow per unit time has been estimated
as $\sim10~{\rm M_\sun~yr^{-1}}$. Thus, in terms of the mass budget,
the Galactic star formation history is controlled by whether the outflow falls
back to the Galactic disk or not, which is determined by the ratio of the thermal
pressure to the CR pressure.
\par
Here we discuss about the boundary conditions at the $z=2$~kpc for the `fall back' solutions
using $T_{\rm bt}\simeq5\times10^6$~K,
$n_{\rm bt}\simeq 5\times10^{-3}~{\rm cm^{-3}}$, $P_{\rm cr,bt}\simeq0.5~{\rm eV~cm^{-3}}$,
and $H_{\rm cr,bt}\simeq34$~kpc as an example (see Table~\ref{tab:bottom three}).
The temperature comparable with the virial temperature for a distance of $\sim100$~kpc
may be required to drive the wind, and it is consistent with the X-ray
observations~\citep[e.g.,][]{nakashima18}. A number density of $\sim10^{-3}\mathchar`-10^{-2}~{\rm cm^{-3}}$
may be nontrivial, though it is supported by numerical simulations~\citep[e.g.,][]{girichidis18}.
If the hot, tenuous gas above the Galactic disk secularly exists
due to the balance of energy injection by supernovae and radiative loss,
$\eta L_{\rm SN}/(\pi R^2z)\sim n^{2}\Lambda$,
where $L_{\rm SN}$ and $\eta$ are the energy injection rate of supernovae
and conversion efficiency, respectively, we obtain
%
%%%%%%%%%%%%%%%%
\begin{eqnarray}
n_{\rm bt}
&\sim&
3\times10^{-3}~{\rm cm^{-3}}
\left( \frac{ \eta       }{ 10^{-2}                        } \right)^{1/2}
\left( \frac{ L_{\rm SN} }{ 10^{42}~{\rm erg~s^{-1} }      } \right)^{1/2}
\nonumber \\
&\times&
\left( \frac{ R          }{ 10~{\rm kpc}                   } \right)^{-1}
\left( \frac{ z          }{  1~{\rm kpc}                   } \right)^{-1/2}
\left( \frac{ \Lambda    }{ 10^{-22}~{\rm erg~cm^3~s^{-1}} } \right)^{-1/2}. 
\end{eqnarray}
%%%%%%%%%%%%%%%%
%
The energy injection rate of the supernovae, $L_{\rm SN}\sim10^{42}~{\rm erg~s^{-1}}$,
corresponds to an event rate of three times per hundred years. It is expected that
roughly a tenth of the kinetic energy of a supernova is converted to the turbulence
in the ISM. Thus, a conversion efficiency of $\eta\sim10^{-2}$ means that roughly a
tenth of the energy of the ISM turbulence is consumed to form
the hot, tenuous gas. The wind may have a velocity comparable to the virial velocity,
$v_{\rm bt}\sim v_{\rm vir}\simeq298~{\rm km~s^{-1}}~(M/10^{12}~{\rm M_\sun})^{1/2}(r/100~{\rm kpc})^{-1/2}$,
which is consistent with our model calculation. The total mass carried by the wind
per unit time can be estimated as
%
%%%%%%%%%%%%%%%%%
\begin{eqnarray}
\dot{M}
&\sim&
\pi R^2 m_{\rm p}n_{\rm bt}v_{\rm vir}
\nonumber \\
&\sim&
6.4~{\rm M_\sun~yr^{-1}}
\left( \frac{ R           }{ 10~{\rm kpc}                 } \right)^2
\nonumber \\
&\times&
\left( \frac{ n_{\rm bt}  }{ 3\times10^{-3}~{\rm cm^{-3}} } \right)
\left( \frac{ v_{\rm vir} }{ 300~{\rm km~s^{-1}}          } \right).
\end{eqnarray}
%%%%%%%%%%%%%%%%%
%
Note that the CR pressure is required to transfer the metals at a height of $\sim100$~kpc
by the wind. When the CR pressure satisfies $P_{\rm g,bt}/P_{\rm cr,bt}\ga1$,
the wind suffers the significant radiative cooling. In this case, the solution cannot be
extended to infinity. This might be related to the possible existence of `fall back' phenomena,
which are interesting as a gas replenishment mechanisms.
Thus, to realize a steady Galactic system, the condition of
%
%%%%%%%%%%%%%%%
\begin{eqnarray}
P_{\rm cr,bt}
&\la&
0.78~{\rm eV~cm^{-3}}
\nonumber \\
&\times&
\left( \frac{ T_{\rm vir} }{ 3\times10^6~{\rm K}          }\right)
\left( \frac{ n_{\rm bt}  }{ 3\times10^{-3}~{\rm cm^{-3}} }\right),
\end{eqnarray}
%%%%%%%%%%%%%%%
%
may be required.
Since the wind carries the CRs, we can estimate an injection power of CRs
in the Galactic disk as
%
%%%%%%%%%%%%%%%
\begin{eqnarray}
L_{\rm cr}
&\sim&
\frac{ P_{\rm cr,bt} }{\gamma_c-1} v_{\rm vir}\pi R^2
\nonumber \\
&\la& 
3.2\times10^{41}~{\rm erg~s^{-1}}
\left( \frac{ P_{\rm cr,bt} }{ 0.78~{\rm eV~cm^{-3}} }\right)
\nonumber \\
&\times&
\left( \frac{ v_{\rm vir}   }{ 300~{\rm km~s^{-1}}   }\right)
\left( \frac{ R             }{ 10~{\rm kpc}          }\right)^2.
\end{eqnarray}
%%%%%%%%%%%%%%%
%
Thus, if the CRs are injected by the supernovae, the wind suffers the significant
radiative cooling. This system can be stable for variations of the star formation rate.
Let us suppose that the star formation rate increases from the current average rate
of a few ${\rm M_\sun}~{\rm yr^{-1}}$. Then, the CR pressure becomes larger than the thermal pressure
due to an increment in the star formation rate (i.e., increment in the event rate of the supernovae),
and the wind can reach at a height of $z>350$~kpc without the significant radiative cooling.
Thus, the Galactic disk secularly loses the gaseous matter, leading to a decrement in the star
formation rate. In contrast, with decreasing the CR pressure, the radiative cooling of wind becomes
significant, which may lead to the `fall back' phenomena and an increment in the star formation rate.
This self-regulation effect by the wind possibly explains the constant star formation
rate averaged by a time scale of $\sim1$~Gyr~\citep[e.g.,][]{haywood16}.
Hence, it is important to study the conversion efficiency $\eta$ and existence of the `fall back' phenomena,
which are not analyzed in this article. We will address them in our future work.
\par
We compare our model with current observations of MW's CGM. \citet{miller15} analyzed
emission line measurements of \ion{O}{8} and \ion{O}{7} from {\it XMM-Newton}/EPIC-MOS spectra,
and gave some constraints for the hot CGM assuming a one-dimensional density structure,
%
%%%%%%%%%%%%%%%
\begin{eqnarray}
n(r)\approx \frac{ n_o r_c{}^{3\beta} }{ r^{3\beta} },
\end{eqnarray}
%%%%%%%%%%%%%%%
%
where $r$ is the galactocentric radius. From the \ion{O}{8} observations, the parameters
$n_o$, $r_c$ and $\beta$ were derived as $n_o r_c{}^{3\beta}=(1.35\mathchar`-1.50)\times10^{-2}$
and $\beta=0.50\mathchar`-0.54$, where we omit to display the $1\sigma$ error, and the range
of parameter values results from optical depth corrections. These constraints
are mainly derived from the \ion{O}{8} measurements. They estimated the total mass of the X-ray emitting gas
as $M_{\rm MB15}=(2.9\mathchar`-3.8)\times{\rm 10^9~M_\sun}$ for $r<50$~kpc and $M_{\rm MB15}=(2.7\mathchar`-4.3)
\times10^{10}~{\rm M_\sun}$ for $r<250$~kpc.\footnote{The analyzed hot X-ray emitting medium was referred as `hot halo' but
the derived length scale is comparable with the CGM we supposed. Therefore, we refer to the `hot halo'
of \citet{miller15} as `CGM' in this article.} In their analysis, the CGM was assumed to
have a constant temperature profile in CIE with fixed $\log(T)=6.3$ and a metallicity of
$Z=0.3Z_\sun$. Although these constraints are based on a different situation from our wind,
the total mass of our wind should has been comparable to the estimated mass given by the
intensity of the emission lines (almost equivalent to the column density). The effects of
the lower metallicity ($Z=0.3Z_\sun$) are discussed later. The order of magnitude estimate of the total mass
may be written as
%
%%%%%%%%%%%%%%%
\begin{eqnarray}
M_{\rm w}
&\sim&
\dot{M}\frac{ z }{ v_{\rm vir} }
\nonumber \\
&\sim& 2.1\times10^9~{\rm M_\sun}
\left( \frac{ R          }{ 10~{\rm kpc}                 } \right)^2
\nonumber \\
&\times&
\left( \frac{ n_{\rm bt} }{ 3\times10^{-3}~{\rm cm^{-3}} } \right)
\left( \frac{ z          }{ 100~{\rm kpc}                } \right).
\end{eqnarray}
%%%%%%%%%%%%%%%
If `fall back' phenomena really occur over the cooling time of $\sim1~{\rm Gyr}$, the expected mass inflow rate
($\sim M_{\rm w}/{\rm 1~Gyr}\sim1~{\rm M_\sun}~{\rm yr}^{-1}$) onto the disk might be sufficiently large in replenishing
the gas of the Galactic disk. From our model calculations, we estimate the statistical average of the total mass as
%
%%%%%%%%%%%%%%%
\begin{eqnarray}
\Delta M_{{\rm w},i}        &=&  2\pi\Delta R_i R_i \int_{z_{\rm bt}}^{z_{\rm tp}} dz \rho(R_i,z), \\
\langle M_{{\rm w}} \rangle &=& \sum_{i=1}^{10} \langle \Delta M_{{\rm w},i} \rangle \times 2, \\
R_i               &=& i~{\rm kpc},~~~\Delta R_i = 1~{\rm kpc}.
\end{eqnarray}
%%%%%%%%%%%%%%%
%
Then, we obtain $\langle M_{\rm w} \rangle\simeq0.51\times10^9~{\rm M_\sun}$ for all solutions and
$\langle M_{\rm w} \rangle\simeq0.66\times10^9~{\rm M_\sun}$ for the `fall back' solutions.
Note that, in our scenario, the observationally constrained mass should include other
components, such as metal-polluted intergalactic medium (IGM) heated by the wind termination shock.
In addition, the mass of the inflow from the CGM to the Galactic disk (i.e., the total mass of the CGM)
should be larger than the mass carried by the wind because a comparable mass is consumed for the star
formation. Thus, the total mass of the wind estimated by our model can still be consistent with the constraints
given by \citet{miller15} in terms of the mass budget that may explain the observed star formation history.
\par
\citet{miller15} also obtained the subsolar metallicity of the hot CGM by combinations
of the emission (\ion{O}{8}) and absorption (\ion{O}{7}) analysis and the pulsar's dispersion
measure toward the Large Magellanic Cloud. Since the lower metallicities reduce the radiative cooling
rate, it might be interesting to investigate how the wind profile may change from our calculations with $Z=Z_\sun$.
%
%%%%%%%%%%%%%%
\begin{figure}[htbp]
\centering
\includegraphics[scale=0.7]{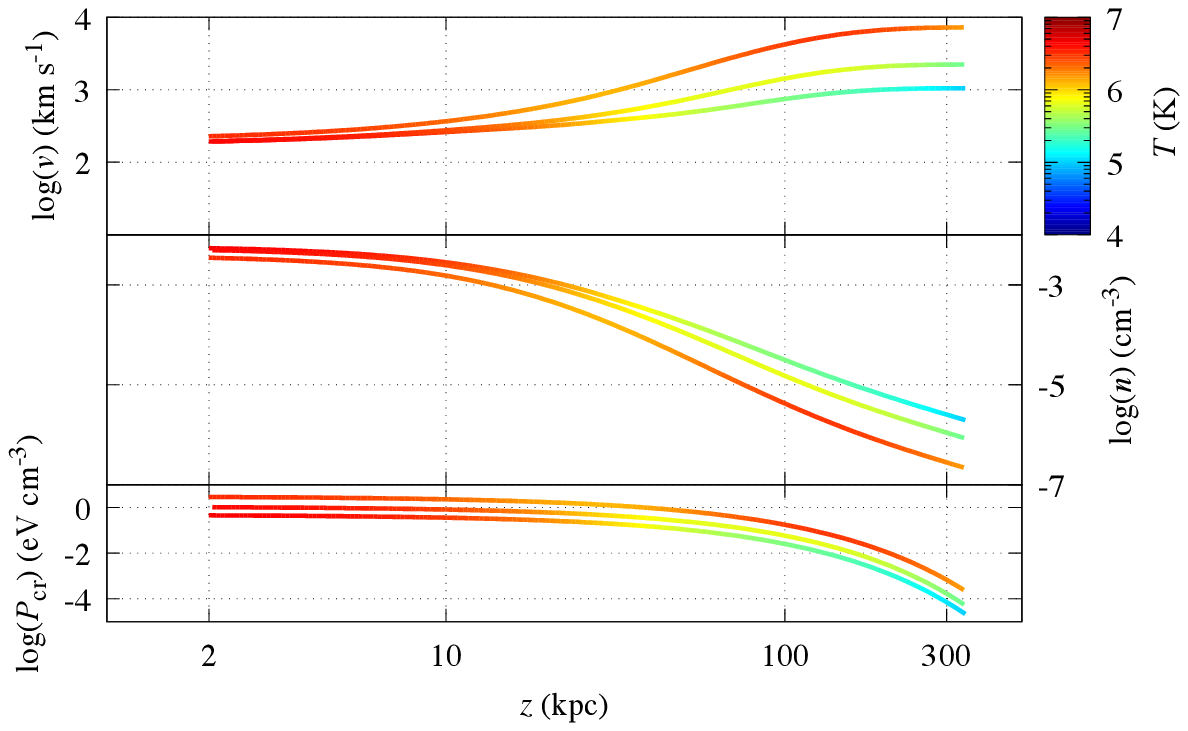}
\includegraphics[scale=0.7]{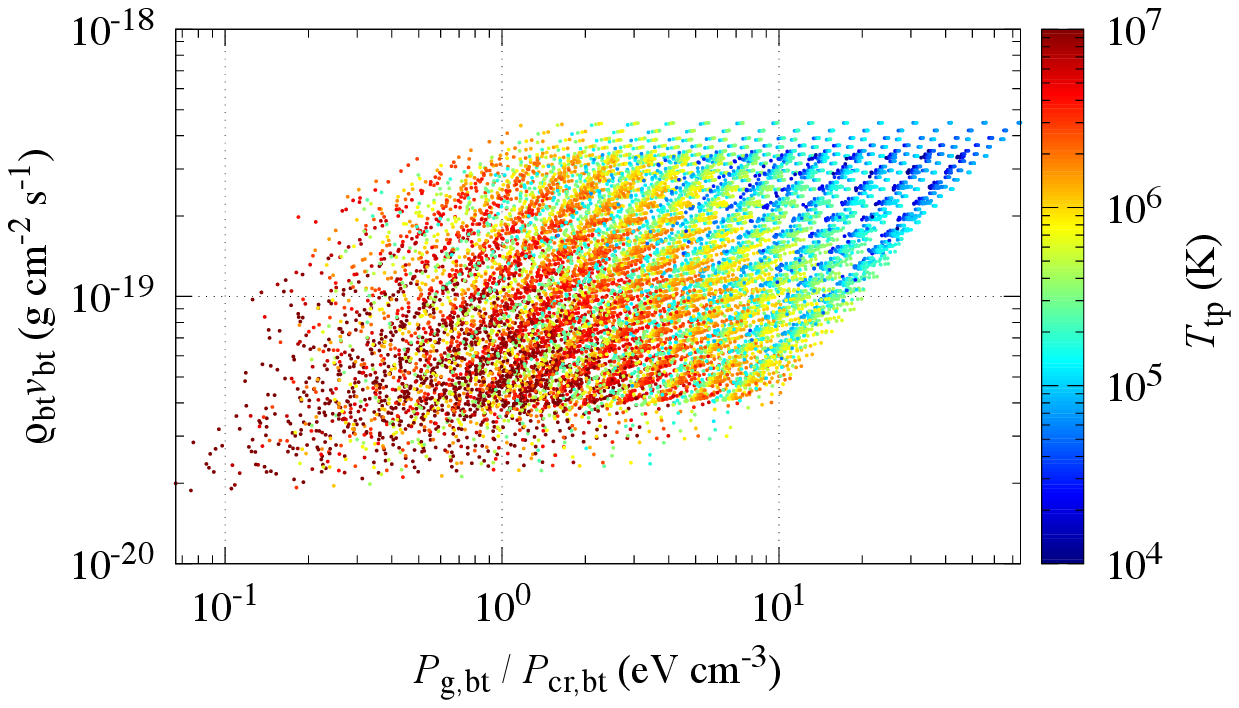}
\caption{
(top panel): The solutions at $R=8$~kpc for
the same parameters as the Fig.~\ref{fig:three solutions}. but the radiative
cooling function is reduced by a factor of 0.1.
The top part shows the velocity, middle part shows the number density,
and bottom part shows the CR pressure. The color indicates the temperature.
(bottom panel): The same as the Figure~\ref{fig:P rhov Ttp} but the radiative cooling
rate is reduced by a factor of 0.1.
The color shows the temperature $T_{\rm tp}$ at the top boundary
($z=z_{\rm tp}$).
}
\label{fig:z=0.1}
\end{figure}
%%%%%%%%%%%%%%
%
The top panel of Figure~\ref{fig:z=0.1} shows the wind solutions with the same
boundary conditions as the case of Figure~\ref{fig:three solutions} but the
radiative cooling rate is reduced by a factor of 0.1. The reduced cooling rate
obviously results in a hotter wind; however, the other profiles of $v$, $n$ and $P_{\rm cr}$
are not so changed. The heating rate due to the CRs is still smaller than the reduced
cooling rate at a lower $z$. Once the wind passes the transonic point, the thermal pressure
becomes less important and the wind is mainly driven by the CR pressure. Thus, for the typical
conditions of the MW we discussed above, the down to $0.1Z_\sun$ does not change our expectations.
This is confirmed by the overall trend of the solutions with the reduced cooling function (the
bottom panel of Figure~\ref{fig:z=0.1}). The result is similar to the case of $Z=Z_\sun$ (Figure~\ref{fig:P rhov Ttp}).
Note that the number of solutions with a reduced cooling rate is 580,014 (127,633 for the `fall back' solutions).
Thus, if we compare the same boundary conditions, the reduced cooling rate results in launching the wind `more easily'
than the case of the solar metallicity. The average total masses are
$\langle M \rangle\simeq0.81\times10^9~{\rm M_\sun}$ for all solutions and $\langle M \rangle
\simeq1.18\times10^9~{\rm M_\sun}$ for the `fall back' solutions. From such insensitive metallicity dependence,
we may regard that our picture of a quasi-steady-star formation in the MW has been continued from $\sim8$~Gyr ago
when the amount of metals was smaller than the current one.
\par
Comparison of the current CR energy density at the Galactic disk with our estimates may be important.
\citet{cerri17} studied the CR propagation in the Galactic disk below a height of $z<2\mathchar`-4$~kpc
to explain the spatial variation of $\gamma$-ray spectral slope observed by Fermi LAT collaboration~\citep{acero16}.
To derive the CR energy density (or pressure) from the $\gamma$-ray observations, we must know the conditions
of the thermal gas including all the components (diffuse hot medium, \ion{H}{1} cloud, molecular cloud, etc.)
in detail because the $\gamma$-ray photons result from the hadronic interaction between the CR protons
and thermal protons ($p_{\rm CR}+p_{\rm ISM}\rightarrow \pi^0 \rightarrow 2\gamma$). However, the conditions of the thermal gas
are not fully understood. Moreover, the propagation of CRs around the molecular cloud is also an unsettled issue in terms of
the effects of the magnetic mirror with a local turbulent field \citep[e.g.,][]{lazarian21}. Thus, it would be better to discuss
the both of the observational estimate and theoretical prediction rather than the observation only.
The propagation model of \citet{cerri17} is based on an anisotropic diffusion coefficient tensor given by pitch-angle scattering;
the diffusion along the guide magnetic field is stronger than the perpendicular one.
The global configuration of the magnetic field controls the spatial distribution of the CR spectral slope and energy
density and is assumed to have a substantial poloidal component. This situation is similar to our model.
Their model does not include the various thermal gas conditions, and the calculated energy density and spectral slope of the CR protons
are directly compared to the estimated values from the observations. This procedure corresponds to CRs that are approximated to
propagate in a diffuse medium with a uniform density structure. The resultant CR energy density of $\sim0.3~{\rm eV~cm^{-3}}$
(equivalently, $P_{\rm cr}\sim0.1~{\rm eV~cm^{-3}}$) can be consistent with the Fermi LAT observations at $R>6$~kpc.\footnote{
The observations and models provide the number density of CR protons with an energy of $E_{\rm cr}>10$~GeV to be $n_{\rm cr}\sim10^{-11}~{\rm cm^{-3}}$.
We estimate the energy density as $10~{\rm GeV} \times 10^{-11}~{\rm cm^{-3}}\times\left(E_{\rm cr}/10~{\rm GeV}\right)^{-0.5}
\sim0.3~{\rm eV~cm^{-3}} \left(E_{\rm cr}/1~{\rm GeV}\right)^{-0.5}$. This estimate may be valid because the spectral slope of CRs is
estimated as $\sim2.2\mathchar`-2.6$ (almost all energy is deposited by CRs with an energy of $E_{\rm cr}\sim1\mathchar`-10$~GeV).}
At $R<6$~kpc, the predicted energy density depleting toward the Galactic center is significantly lower than that estimated from the
observations $\sim0.3\mathchar`-1~{\rm eV~cm^{-3}}$ ($P_{\rm cr}\sim0.1\mathchar`-0.3~{\rm eV~cm^{-3}}$). These estimates satisfy
the inequality $P_{\rm cr,bt}\la0.78~{\rm eV~cm^{-3}}$ we discussed above. Note that the CR pressure of $\sim0.1\mathchar`-0.3~{\rm eV~cm^{-3}}$
is sufficiently large in launching the wind depending on the thermal gas conditions. Thus, the current state of the MW implies that
the cold wind is driven, eventually falling back to the disk in the future.
\par
Observations of external galaxies show that the CGM consists of not
only highly ionized species like \ion{O}{6} but also lower-ionized
species like \ion{H}{1}, \ion{C}{2}, \ion{Mg}{2}, and so on
\citep[][and references therein]{tumlinson17}.
The absorption lines of these lower-ionized species are also observed
at a distance of $\sim100$~kpc from the host galaxy.
Our solutions imply that the outflow can transfer such lower-ionized
species to a height of $z\sim100$~kpc if the radiative cooling is efficient.
The condition for efficient radiative cooling is $P_{\rm g,bt}/P_{\rm cr,bt}\ga1$.
Note that the ionization state is affected by photoionization which is omitted
in this paper. Thus, we have to pay attention whether the \ion{O}{6} absorption line
really indicates the existence of hot gas~\citep[see][]{breitschwerdt99}.
On the other hand, photoionization reduces the number of lower-ionized species.
Thus, the existence of the lower ionized species indicates that the temperature
of gas is low or the condensation of gas occurs to shield itself from the photoionizing
photons. In any case, an efficient cooling process is required.
We will study the ionization state and condensation of gas at a height of $\sim100$~kpc
in future work.
\par
We have assumed that the outflow travels along the vertical direction $z$
so that the required energy is approximately minimum. This condition, however,
is strictly given by $\bm{v}\propto\bm{\nabla}\Phi$. Thus, the outflow may have
at least two-dimensional spatial structure, like a biconical structure, in reality.
It is reported by numerical simulation that the CR pressure can affect the spatial structure
of wind~\citep{hopkins18}. The far-ultraviolet observation of NGC~3079 implies an X-shape
wind~\citep{hodges-kluck20}. Thus, it would be worth to investigating the relation between
the wind condition and its morphology.
The wind morphology may depend on how the diffusion coefficient is assumed. The multidimensional
diffusion coefficient (i.e., diffusion tensor) is also actively discussed issue in the CR transport
literature~\citep[e.g.,][, and references therein]{cerri17}. \citet{zirakashvili96} studied an axially
symmetric wind considering the Galactic disk rotation and introduced an effective CR adiabatic index
$\gamma_{\rm eff}$ as a possibly useful method,
%
%%%%%%%%%%%%%%%
\begin{eqnarray}
\frac{\gamma_{\rm eff} }{ \gamma_{\rm eff} - 1 }
=\frac{ \gamma_c }{ \gamma_c - 1 }
-\frac{ \kappa }{ \left(\gamma_c-1\right)\left( v+V_{\rm A}\right) }
\frac{1}{P_{\rm cr}}\frac{ dP_{\rm cr} }{dz},
\end{eqnarray}
%%%%%%%%%%%%%%%
%
so that the CR pressure can be expressed as $P_{\rm cr}\left[A(v+V_{\rm A})\right]^{-\gamma_{\rm eff}}=~{\rm const}$.
It may simplify the analysis of the transonic `surface' of the multidimenstional wind and the systematic
parameter study of the diffusion coefficient.
We hope to extend our model to a multidimensional one in our future work.

\acknowledgments
We thank K. Masai for useful discussions.
We are grateful to the anonymous referee, for his/her comments that further improved the paper.
This work is supported by JSPS Grants-in-Aid for Scientific Research Nos. 20J01086 (JS),
16H02160, 18H05436, and 18H05437 (SI).

\software{Cloudy \citep{ferland17}}

\appendix
The expressions of ${\cal N}_z$ and ${\cal N}_v$ are, respectively,
%
%%%%%%%%%%%%%%
\begin{eqnarray}
&& {\cal N}_z = z\frac{A'}{A}
\frac{ \partial }{ \partial z }\left( C_{\rm g}{}^2+C_{\rm w}{}^2-V_{\rm g}{}^2 \right),
\\
&& \frac{ \partial C_{\rm g}{}^2 }{  \partial z }
= - \left( \gamma_{\rm g} - 1 \right)
\left[
  C_{\rm g}{}^2 \frac{A'}{A}
  + \frac{ \gamma_{\rm g} }{ \rho v }
    \left( V_{\rm A} \frac{ dP_{\rm cr} }{ dz }
           -n^2\Lambda \right)
\right],
\nonumber \\
&& \frac{ \partial C_{\rm w}{}^2 }{ \partial z }
= - \frac{ 3M_{\rm A}{}^2-2M_{\rm A}-3 }{ 2(3M_{\rm A}+1)(M_{\rm A}+1) }C_{\rm w}{}^2\frac{A'}{A},
\nonumber \\
&& \frac{ \partial V_{\rm g}{}^2 }{ \partial z }
= \left( 1 - \frac{ Z_0{}^2 + z^2 }{ 2z^2 } \right)
  \left[ \frac{ d\Phi }{ dz }
       - \left( \frac{ \gamma_{\rm g}-1 }{ M_{\rm A} } - 1 \right)  \right]
       + \frac{A'}{A}\left[ \frac{ d^2\Phi }{ dz^2 }
                      - \frac{ \partial {\cal F} }{ \partial z } \right],
\nonumber \\
&& \frac{ \partial {\cal F} }{ \partial z }
= -\frac{ (\gamma_{\rm g}-1) n^2 }{ \rho v }
\left[ 
         \frac{ \gamma_{\rm g}-1 }{ k_{\rm B} }
         \left\{ \frac{n\Lambda}{v} + k_{\rm B}T\frac{A'}{A} \right\}
         \frac{ \partial \Lambda }{ \partial T }
       +\Lambda\frac{A'}{A}
\right]
- \frac{ \gamma_{\rm g}-1 }{ 2M_{\rm A} }\frac{A'}{A}\frac{ P_{\rm cr}' }{\rho}
\nonumber \\
&&~~~~~~~
+ \left( \frac{ \gamma_{\rm g}-1 }{ M_{\rm A} } - 1 \right)
\left\{ \left(
        \frac{ d\ln P_{\rm cr} }{ dz }+\frac{A'}{A}
        \right)
        \frac{ 1 }{ \rho }\frac{ dP_{\rm cr} }{ dz }
          + \frac{ P_{\rm cr}            }{ \rho        }
            \frac{      \partial         }{  \partial z }
            \frac{ d\ln P_{\rm cr}       }{ dz          }
\right\},
\nonumber \\
&& \frac{  \partial         }{  \partial z }
   \frac{ d\ln P_{\rm cr}   }{ dz          }
= -\left( \frac{ d\ln P_{\rm cr} }{ dz       } \right)^2 
+  \left( \frac{ v + V_{\rm A}   }{ \kappa   }
        - \frac{ A'              }{ A        }
        - \frac{ \kappa'         }{ \kappa   } \right) 
          \frac{ d\ln P_{\rm cr} }{ dz       }
        + \frac{ \gamma_c ( v + V_{\rm A}/2) }{ \kappa } \frac{ A' }{ A },
\nonumber
\end{eqnarray}
and
\begin{eqnarray}
{\cal N}_v = z\frac{A'}{A}
\left[
\frac{ C_{\rm g}{}^2 }{ v }
+ \left\{
   1 + \frac{ M_{\rm A} }{ 2(M_{\rm A}+1)(3M_{\rm A}+1) }
  \right\} \frac{ C_{\rm w}{}^2 }{ v }
- \frac{A'}{A}
\left\{
2(\gamma_{\rm g}-1)\frac{ n^2\Lambda }{ \rho v^2 }
- \left( \frac{ \gamma_{\rm g}-1 }{ 2M_{\rm A} } - 1 \right)
\frac{1}{\rho v}\frac{ dP_{\rm cr} }{ dz }
\right\}
\right].
\end{eqnarray}
%%%%%%%%%%%%%%
%


%\vspace{5mm}
%\facilities{HST(STIS), Swift(XRT and UVOT), AAVSO, CTIO:1.3m,


%\software{astropy \citep{2013A&A...558A..33A},  
%          Cloudy \citep{2013RMxAA..49..137F}, 
%          SExtractor \citep{1996A&AS..117..393B}
%          }



\bibliography{apj_sjsi21}{}

\begin{thebibliography}{}
\expandafter\ifx\csname natexlab\endcsname\relax\def\natexlab#1{#1}\fi
\providecommand{\url}[1]{\href{#1}{#1}}
\providecommand{\dodoi}[1]{doi:~\href{http://doi.org/#1}{\nolinkurl{#1}}}
\providecommand{\doeprint}[1]{\href{http://ascl.net/#1}{\nolinkurl{http://ascl.net/#1}}}
\providecommand{\doarXiv}[1]{\href{https://arxiv.org/abs/#1}{\nolinkurl{https://arxiv.org/abs/#1}}}

\bibitem[{{Acero} {et~al.}(2016){Acero}, {Ackermann}, {Ajello}, {Albert},
  {Baldini}, {Ballet}, {Barbiellini}, {Bastieri}, {Bellazzini}, {Bissaldi},
  {Bloom}, {Bonino}, {Bottacini}, {Brandt}, {Bregeon}, {Bruel}, {Buehler},
  {Buson}, {Caliandro}, {Cameron}, {Caragiulo}, {Caraveo}, {Casandjian},
  {Cavazzuti}, {Cecchi}, {Charles}, {Chekhtman}, {Chiang}, {Chiaro}, {Ciprini},
  {Claus}, {Cohen-Tanugi}, {Conrad}, {Cuoco}, {Cutini}, {D'Ammando}, {de
  Angelis}, {de Palma}, {Desiante}, {Digel}, {Di Venere}, {Drell}, {Favuzzi},
  {Fegan}, {Ferrara}, {Focke}, {Franckowiak}, {Funk}, {Fusco}, {Gargano},
  {Gasparrini}, {Giglietto}, {Giordano}, {Giroletti}, {Glanzman}, {Godfrey},
  {Grenier}, {Guiriec}, {Hadasch}, {Harding}, {Hayashi}, {Hays}, {Hewitt},
  {Hill}, {Horan}, {Hou}, {Jogler}, {J{\'o}hannesson}, {Kamae}, {Kuss},
  {Landriu}, {Larsson}, {Latronico}, {Li}, {Li}, {Longo}, {Loparco},
  {Lovellette}, {Lubrano}, {Maldera}, {Malyshev}, {Manfreda}, {Martin},
  {Mayer}, {Mazziotta}, {McEnery}, {Michelson}, {Mirabal}, {Mizuno}, {Monzani},
  {Morselli}, {Nuss}, {Ohsugi}, {Omodei}, {Orienti}, {Orlando}, {Ormes},
  {Paneque}, {Pesce-Rollins}, {Piron}, {Pivato}, {Rain{\`o}}, {Rando},
  {Razzano}, {Razzaque}, {Reimer}, {Reimer}, {Remy}, {Renault},
  {S{\'a}nchez-Conde}, {Schaal}, {Schulz}, {Sgr{\`o}}, {Siskind}, {Spada},
  {Spandre}, {Spinelli}, {Strong}, {Suson}, {Tajima}, {Takahashi}, {Thayer},
  {Thompson}, {Tibaldo}, {Tinivella}, {Torres}, {Tosti}, {Troja}, {Vianello},
  {Werner}, {Wood}, {Wood}, {Zaharijas}, \& {Zimmer}}]{acero16}
{Acero}, F., {Ackermann}, M., {Ajello}, M., {et~al.} 2016, \apjs, 223, 26,
  \dodoi{10.3847/0067-0049/223/2/26}

\bibitem[{{Achterberg}(1981{\natexlab{a}})}]{achterberg81a}
{Achterberg}, A. 1981{\natexlab{a}}, \aap, 98, 195

\bibitem[{{Achterberg}(1981{\natexlab{b}})}]{achterberg81b}
---. 1981{\natexlab{b}}, \aap, 98, 161

\bibitem[{{Altun} {et~al.}(2007){Altun}, {Yumak}, {Yavuz}, {Badnell}, {Loch},
  \& {Pindzola}}]{altun07}
{Altun}, Z., {Yumak}, A., {Yavuz}, I., {et~al.} 2007, \aap, 474, 1051,
  \dodoi{10.1051/0004-6361:20078238}

\bibitem[{{Arnaud} \& {Rothenflug}(1985)}]{arnaud85}
{Arnaud}, M., \& {Rothenflug}, R. 1985, \aaps, 60, 425

\bibitem[{{Asplund} {et~al.}(2009){Asplund}, {Grevesse}, {Sauval}, \&
  {Scott}}]{asplund09}
{Asplund}, M., {Grevesse}, N., {Sauval}, A.~J., \& {Scott}, P. 2009, \araa, 47,
  481, \dodoi{10.1146/annurev.astro.46.060407.145222}

\bibitem[{{Boulares} \& {Cox}(1990)}]{boulares90}
{Boulares}, A., \& {Cox}, D.~P. 1990, \apj, 365, 544, \dodoi{10.1086/169509}

\bibitem[{{Bregman} \& {Lloyd-Davies}(2007)}]{bregman07}
{Bregman}, J.~N., \& {Lloyd-Davies}, E.~J. 2007, \apj, 669, 990,
  \dodoi{10.1086/521321}

\bibitem[{{Breitschwerdt} {et~al.}(1991){Breitschwerdt}, {McKenzie}, \&
  {Voelk}}]{breitschwerdt91}
{Breitschwerdt}, D., {McKenzie}, J.~F., \& {Voelk}, H.~J. 1991, \aap, 245, 79

\bibitem[{{Breitschwerdt} \& {Schmutzler}(1999)}]{breitschwerdt99}
{Breitschwerdt}, D., \& {Schmutzler}, T. 1999, \aap, 347, 650.
\newblock \doarXiv{astro-ph/9902268}

\bibitem[{{Cerri} {et~al.}(2017){Cerri}, {Gaggero}, {Vittino}, {Evoli}, \&
  {Grasso}}]{cerri17}
{Cerri}, S.~S., {Gaggero}, D., {Vittino}, A., {Evoli}, C., \& {Grasso}, D.
  2017, \jcap, 2017, 019, \dodoi{10.1088/1475-7516/2017/10/019}

\bibitem[{{Ferland} {et~al.}(2017){Ferland}, {Chatzikos}, {Guzm{\'a}n},
  {Lykins}, {van Hoof}, {Williams}, {Abel}, {Badnell}, {Keenan}, {Porter}, \&
  {Stancil}}]{ferland17}
{Ferland}, G.~J., {Chatzikos}, M., {Guzm{\'a}n}, F., {et~al.} 2017, \rmxaa, 53,
  385.
\newblock \doarXiv{1705.10877}

\bibitem[{{Ferri{\`e}re}(2001)}]{ferriere01}
{Ferri{\`e}re}, K.~M. 2001, Reviews of Modern Physics, 73, 1031,
  \dodoi{10.1103/RevModPhys.73.1031}

\bibitem[{{Gabici} {et~al.}(2019){Gabici}, {Evoli}, {Gaggero}, {Lipari},
  {Mertsch}, {Orlando}, {Strong}, \& {Vittino}}]{gabici19}
{Gabici}, S., {Evoli}, C., {Gaggero}, D., {et~al.} 2019, International Journal
  of Modern Physics D, 28, 1930022, \dodoi{10.1142/S0218271819300222}

\bibitem[{{Ginzburg} \& {Syrovatskii}(1964)}]{ginzburg64}
{Ginzburg}, V.~L., \& {Syrovatskii}, S.~I. 1964, {The Origin of Cosmic Rays}

\bibitem[{{Girichidis} {et~al.}(2018){Girichidis}, {Naab}, {Hanasz}, \&
  {Walch}}]{girichidis18}
{Girichidis}, P., {Naab}, T., {Hanasz}, M., \& {Walch}, S. 2018, \mnras, 479,
  3042, \dodoi{10.1093/mnras/sty1653}

\bibitem[{{Gnat}(2017)}]{gnat17}
{Gnat}, O. 2017, \apjs, 228, 11, \dodoi{10.3847/1538-4365/228/2/11}

\bibitem[{{Gronenschild} \& {Mewe}(1978)}]{gronen78}
{Gronenschild}, E.~H.~B.~M., \& {Mewe}, R. 1978, \aaps, 32, 283

\bibitem[{{Hahn} {et~al.}(2014){Hahn}, {Badnell}, {Grieser}, {Krantz},
  {Lestinsky}, {M{\"u}ller}, {Novotn{\'y}}, {Repnow}, {Schippers}, {Wolf}, \&
  {Savin}}]{hahn14}
{Hahn}, M., {Badnell}, N.~R., {Grieser}, M., {et~al.} 2014, \apj, 788, 46,
  \dodoi{10.1088/0004-637X/788/1/46}

\bibitem[{{Hayakawa} {et~al.}(1958){Hayakawa}, {Ito}, \&
  {Terashima}}]{hayakawa58}
{Hayakawa}, S., {Ito}, K., \& {Terashima}, Y. 1958, Progress of Theoretical
  Physics Supplement, 6, 1, \dodoi{10.1143/PTPS.6.1}

\bibitem[{{Haywood} {et~al.}(2016){Haywood}, {Lehnert}, {Di Matteo}, {Snaith},
  {Schultheis}, {Katz}, \& {G{\'o}mez}}]{haywood16}
{Haywood}, M., {Lehnert}, M.~D., {Di Matteo}, P., {et~al.} 2016, \aap, 589,
  A66, \dodoi{10.1051/0004-6361/201527567}

\bibitem[{{Hodges-Kluck} {et~al.}(2020){Hodges-Kluck}, {Yukita}, {Tanner},
  {Ptak}, {Bregman}, \& {Li}}]{hodges-kluck20}
{Hodges-Kluck}, E.~J., {Yukita}, M., {Tanner}, R., {et~al.} 2020, \apj, 903,
  35, \dodoi{10.3847/1538-4357/abb884}

\bibitem[{{Hopkins} {et~al.}(2018){Hopkins}, {Wetzel}, {Kere{\v{s}}},
  {Faucher-Gigu{\`e}re}, {Quataert}, {Boylan-Kolchin}, {Murray}, {Hayward},
  {Garrison-Kimmel}, {Hummels}, {Feldmann}, {Torrey}, {Ma},
  {Angl{\'e}s-Alc{\'a}zar}, {Su}, {Orr}, {Schmitz}, {Escala}, {Sanderson},
  {Grudi{\'c}}, {Hafen}, {Kim}, {Fitts}, {Bullock}, {Wheeler}, {Chan},
  {Elbert}, \& {Narayanan}}]{hopkins18}
{Hopkins}, P.~F., {Wetzel}, A., {Kere{\v{s}}}, D., {et~al.} 2018, \mnras, 480,
  800, \dodoi{10.1093/mnras/sty1690}

\bibitem[{{Inutsuka} {et~al.}(2015){Inutsuka}, {Inoue}, {Iwasaki}, {Stone},
  {Suzuki}, {Tsukamoto}, \& {Takamoto}}]{inutsuka15}
{Inutsuka}, S.~i., {Inoue}, T., {Iwasaki}, K., {et~al.} 2015, in Astronomical
  Society of the Pacific Conference Series, Vol. 498, Numerical Modeling of
  Space Plasma Flows ASTRONUM-2014, ed. N.~V. {Pogorelov}, E.~{Audit}, \& G.~P.
  {Zank}, 75

\bibitem[{{Ipavich}(1975)}]{ipavich75}
{Ipavich}, F.~M. 1975, \apj, 196, 107, \dodoi{10.1086/153397}

\bibitem[{{Janev} \& {Smith}(1993)}]{janev93}
{Janev}, R.~K., \& {Smith}, J.~J. 1993, {Cross Sections for Collision Processes
  of Hydrogen Atoms with Electrons, Protons and Multiply Charged Ions}, 192

\bibitem[{{Jokipii}(1966)}]{jokipii66}
{Jokipii}, J.~R. 1966, \apj, 146, 480, \dodoi{10.1086/148912}

\bibitem[{{Kennicutt} \& {Evans}(2012)}]{kennicutt12}
{Kennicutt}, R.~C., \& {Evans}, N.~J. 2012, \araa, 50, 531,
  \dodoi{10.1146/annurev-astro-081811-125610}

\bibitem[{{Kotelnikov} \& {Milstein}(2019)}]{kotelnikov19}
{Kotelnikov}, I.~A., \& {Milstein}, A.~I. 2019, \physscr, 94, 055403,
  \dodoi{10.1088/1402-4896/ab060a}

\bibitem[{{Kulsrud} \& {Pearce}(1969)}]{kulsrud69}
{Kulsrud}, R., \& {Pearce}, W.~P. 1969, \apj, 156, 445, \dodoi{10.1086/149981}

\bibitem[{{Kulsrud}(2005)}]{kulsrud05}
{Kulsrud}, R.~M. 2005, {Plasma physics for astrophysics}

\bibitem[{{Lazarian} \& {Xu}(2021)}]{lazarian21}
{Lazarian}, A., \& {Xu}, S. 2021, arXiv e-prints, arXiv:2106.08362.
\newblock \doarXiv{2106.08362}

\bibitem[{{Lee} \& {V{\"o}lk}(1973)}]{lee73}
{Lee}, M.~A., \& {V{\"o}lk}, H.~J. 1973, \apss, 24, 31,
  \dodoi{10.1007/BF00648673}

\bibitem[{{Lennon} {et~al.}(1988){Lennon}, {Bell}, {Gilbody}, {Hughes},
  {Kingston}, {Murray}, \& {Smith}}]{lennon88}
{Lennon}, M.~A., {Bell}, K.~L., {Gilbody}, H.~B., {et~al.} 1988, Journal of
  Physical and Chemical Reference Data, 17, 1285, \dodoi{10.1063/1.555809}

\bibitem[{{Lerche}(1966)}]{lerche66}
{Lerche}, I. 1966, Physics of Fluids, 9, 1073, \dodoi{10.1063/1.1761804}

\bibitem[{{Lerche}(1967)}]{lerche67}
---. 1967, \apj, 147, 689, \dodoi{10.1086/149045}

\bibitem[{{Lestinsky} {et~al.}(2009){Lestinsky}, {Badnell}, {Bernhardt},
  {Grieser}, {Hoffmann}, {Luki{\'c}}, {M{\"u}ller}, {Orlov}, {Repnow}, {Savin},
  {Schmidt}, {Schnell}, {Schippers}, {Wolf}, \& {Yu}}]{lestinsky09}
{Lestinsky}, M., {Badnell}, N.~R., {Bernhardt}, D., {et~al.} 2009, \apj, 698,
  648, \dodoi{10.1088/0004-637X/698/1/648}

\bibitem[{{Lockman}(1984)}]{lockman84}
{Lockman}, F.~J. 1984, \apj, 283, 90, \dodoi{10.1086/162277}

\bibitem[{{McKee} \& {Ostriker}(1977)}]{mckee77}
{McKee}, C.~F., \& {Ostriker}, J.~P. 1977, \apj, 218, 148,
  \dodoi{10.1086/155667}

\bibitem[{{Mewe}(1972)}]{mewe72}
{Mewe}, R. 1972, \aap, 20, 215

\bibitem[{{Mewe} {et~al.}(1986){Mewe}, {Lemen}, \& {van den Oord}}]{mewe86}
{Mewe}, R., {Lemen}, J.~R., \& {van den Oord}, G.~H.~J. 1986, \aaps, 65, 511

\bibitem[{{Mewe} {et~al.}(1980{\natexlab{a}}){Mewe}, {Schrijver}, \&
  {Sylwester}}]{mewe80a}
{Mewe}, R., {Schrijver}, J., \& {Sylwester}, J. 1980{\natexlab{a}}, \aaps, 40,
  323

\bibitem[{{Mewe} {et~al.}(1980{\natexlab{b}}){Mewe}, {Schrijver}, \&
  {Sylwester}}]{mewe80b}
---. 1980{\natexlab{b}}, \aap, 87, 55

\bibitem[{{Miller} \& {Bregman}(2015)}]{miller15}
{Miller}, M.~J., \& {Bregman}, J.~N. 2015, \apj, 800, 14,
  \dodoi{10.1088/0004-637X/800/1/14}

\bibitem[{{Mitnik} \& {Badnell}(2004)}]{mitnik04}
{Mitnik}, D.~M., \& {Badnell}, N.~R. 2004, \aap, 425, 1153,
  \dodoi{10.1051/0004-6361:20041297}

\bibitem[{{Miyamoto} \& {Nagai}(1975)}]{miyamoto75}
{Miyamoto}, M., \& {Nagai}, R. 1975, \pasj, 27, 533

\bibitem[{{Murakami} {et~al.}(2006){Murakami}, {Kato}, {Kato}, {Safronova},
  {Cowan}, \& {Ralchenko}}]{murakami06}
{Murakami}, I., {Kato}, T., {Kato}, D., {et~al.} 2006, Journal of Physics B
  Atomic Molecular Physics, 39, 2917, \dodoi{10.1088/0953-4075/39/14/001}

\bibitem[{{Nahar}(1995)}]{nahar95}
{Nahar}, S.~N. 1995, \apjs, 101, 423, \dodoi{10.1086/192248}

\bibitem[{{Nahar}(1998)}]{nahar98}
---. 1998, \pra, 58, 3766, \dodoi{10.1103/PhysRevA.58.3766}

\bibitem[{{Nahar}(2000)}]{nahar00}
---. 2000, \apjs, 126, 537, \dodoi{10.1086/313307}

\bibitem[{{Nahar}(2006)}]{nahar06}
---. 2006, \apjs, 164, 280, \dodoi{10.1086/501503}

\bibitem[{{Nahar} \& {Pradhan}(1997)}]{nahar97}
{Nahar}, S.~N., \& {Pradhan}, A.~K. 1997, \apjs, 111, 339,
  \dodoi{10.1086/313013}

\bibitem[{{Nahar} \& {Pradhan}(1999)}]{nahar99}
---. 1999, \aaps, 135, 347, \dodoi{10.1051/aas:1999447}

\bibitem[{{Nahar} {et~al.}(2001){Nahar}, {Pradhan}, \& {Zhang}}]{nahar01}
{Nahar}, S.~N., {Pradhan}, A.~K., \& {Zhang}, H.~L. 2001, \apjs, 133, 255,
  \dodoi{10.1086/319187}

\bibitem[{{Nakashima} {et~al.}(2018){Nakashima}, {Inoue}, {Yamasaki}, {Sofue},
  {Kataoka}, \& {Sakai}}]{nakashima18}
{Nakashima}, S., {Inoue}, Y., {Yamasaki}, N., {et~al.} 2018, \apj, 862, 34,
  \dodoi{10.3847/1538-4357/aacceb}

\bibitem[{{Navarro} {et~al.}(1996){Navarro}, {Frenk}, \& {White}}]{nfw96}
{Navarro}, J.~F., {Frenk}, C.~S., \& {White}, S. D.~M. 1996, \apj, 462, 563,
  \dodoi{10.1086/177173}

\bibitem[{{Novotn{\'y}} {et~al.}(2012){Novotn{\'y}}, {Badnell}, {Bernhardt},
  {Grieser}, {Hahn}, {Krantz}, {Lestinsky}, {M{\"u}ller}, {Repnow},
  {Schippers}, {Wolf}, \& {Savin}}]{novotny12}
{Novotn{\'y}}, O., {Badnell}, N.~R., {Bernhardt}, D., {et~al.} 2012, \apj, 753,
  57, \dodoi{10.1088/0004-637X/753/1/57}

\bibitem[{{Osterbrock} \& {Ferland}(2006)}]{Osterbrock06}
{Osterbrock}, D.~E., \& {Ferland}, G.~J. 2006, {Astrophysics of gaseous nebulae
  and active galactic nuclei}

\bibitem[{{Recchia} {et~al.}(2016){Recchia}, {Blasi}, \& {Morlino}}]{recchia16}
{Recchia}, S., {Blasi}, P., \& {Morlino}, G. 2016, \mnras, 462, L88,
  \dodoi{10.1093/mnrasl/slw136}

\bibitem[{{Savin} {et~al.}(2002){Savin}, {Behar}, {Kahn}, {Gwinner}, {Saghiri},
  {Schmitt}, {Grieser}, {Repnow}, {Schwalm}, {Wolf}, {Bartsch}, {M{\"u}ller},
  {Schippers}, {Badnell}, {Chen}, \& {Gorczyca}}]{savin02}
{Savin}, D.~W., {Behar}, E., {Kahn}, S.~M., {et~al.} 2002, \apjs, 138, 337,
  \dodoi{10.1086/323388}

\bibitem[{{Shapiro} \& {Field}(1976)}]{shapiro76}
{Shapiro}, P.~R., \& {Field}, G.~B. 1976, \apj, 205, 762,
  \dodoi{10.1086/154332}

\bibitem[{{Sofue}(2012)}]{sofue12}
{Sofue}, Y. 2012, \pasj, 64, 75, \dodoi{10.1093/pasj/64.4.75}

\bibitem[{{Tumlinson} {et~al.}(2017){Tumlinson}, {Peeples}, \&
  {Werk}}]{tumlinson17}
{Tumlinson}, J., {Peeples}, M.~S., \& {Werk}, J.~K. 2017, \araa, 55, 389,
  \dodoi{10.1146/annurev-astro-091916-055240}

\bibitem[{{Volk} \& {McKenzie}(1981)}]{volk81}
{Volk}, H.~J., \& {McKenzie}, J.~F. 1981, in International Cosmic Ray
  Conference, Vol.~9, International Cosmic Ray Conference, 246--249

\bibitem[{{Wentzel}(1968)}]{wetzel68}
{Wentzel}, D.~G. 1968, \apj, 152, 987, \dodoi{10.1086/149611}

\bibitem[{{Zatsarinny} {et~al.}(2006){Zatsarinny}, {Gorczyca}, {Fu}, {Korista},
  {Badnell}, \& {Savin}}]{zatsarinny06}
{Zatsarinny}, O., {Gorczyca}, T.~W., {Fu}, J., {et~al.} 2006, \aap, 447, 379,
  \dodoi{10.1051/0004-6361:20053737}

\bibitem[{{Zatsarinny} {et~al.}(2003){Zatsarinny}, {Gorczyca}, {Korista},
  {Badnell}, \& {Savin}}]{zatsarinny03}
{Zatsarinny}, O., {Gorczyca}, T.~W., {Korista}, K.~T., {Badnell}, N.~R., \&
  {Savin}, D.~W. 2003, \aap, 412, 587, \dodoi{10.1051/0004-6361:20031462}

\bibitem[{{Zatsarinny} {et~al.}(2004){Zatsarinny}, {Gorczyca}, {Korista},
  {Badnell}, \& {Savin}}]{zatsarinny04}
---. 2004, \aap, 417, 1173, \dodoi{10.1051/0004-6361:20034174}

\bibitem[{{Zirakashvili} {et~al.}(1996){Zirakashvili}, {Breitschwerdt},
  {Ptuskin}, \& {Voelk}}]{zirakashvili96}
{Zirakashvili}, V.~N., {Breitschwerdt}, D., {Ptuskin}, V.~S., \& {Voelk}, H.~J.
  1996, \aap, 311, 113

\end{thebibliography}
\bibliographystyle{aasjournal}

\end{document}